\documentclass[english,american]{revtex4-2}
\usepackage[T1]{fontenc}
\usepackage[utf8]{inputenc}
\usepackage{xcolor}
\usepackage{amsmath}
\usepackage{amssymb}
\usepackage{graphicx}
\usepackage{esint}
\PassOptionsToPackage{normalem}{ulem}
\usepackage{ulem}
\usepackage{hyperref}
\hypersetup{
    colorlinks=true,
    linkcolor=blue,
    filecolor=magenta,      
    urlcolor=blue,
    citecolor=blue
    }

\usepackage{chngcntr}

\usepackage{titlesec}



\usepackage{babel}
\begin{document}
\selectlanguage{english}%
\global\long\def\ket#1{\left|#1\right\rangle }%

\global\long\def\bra#1{\left\langle #1\right|}%

\global\long\def\ketL#1{\left.\left|#1\right\rangle \right\rangle }%

\global\long\def\braL#1{\left\langle \left\langle #1\right|\right.}%

\global\long\def\braket#1#2{\left\langle #1\left|#2\right.\right\rangle }%

\global\long\def\ketbra#1#2{\left|#1\right\rangle \left\langle #2\right|}%

\global\long\def\braOket#1#2#3{\left\langle #1\left|#2\right|#3\right\rangle }%

\global\long\def\mc#1{\mathcal{#1}}%

\global\long\def\nrm#1{\left\Vert #1\right\Vert }%

\title{Layered KIK quantum error mitigation for dynamic circuits}
\author{Ben Bar}
\email{ben.bar@mail.huji.ac.il}
\author{Jader P. Santos}
\email{jader.pereira.santos@gmail.com}
\author{Raam Uzdin}
\email{raam@mail.huji.ac.il}

\affiliation{The Hebrew University of Jerusalem, Jerusalem, Israel 9190401}

\begin{abstract}
Quantum Error Mitigation is essential for enhancing the reliability of quantum computing experiments. The adaptive KIK error mitigation method has demonstrated significant advantages, including resilience to temporal noise drifts, applicability to non-Clifford gates, and guaranteed performance bounds. However, its reliance on global noise amplification introduces limitations, such as incompatibility with mid-circuit measurements and dynamic circuits, as well as small residual errors due to unaccounted high-order Magnus noise terms. In this work, we propose a layer-based noise amplification approach that overcomes these challenges without incurring additional overhead or experimental complexity. Since the Layered KIK method is inherently compatible with mid-circuit measurements, it enables seamless integration with quantum error correction codes. This synergy allows error correction to address dominant noise mechanisms, while the Layered KIK suppresses residual errors arising from leakage and correlated noise sources. Similarly, for reducing sampling costs, Layered KIK can be combined with complementary mitigation methods for providing drift resilience and broadening the range of addressable errors.
\end{abstract}

\maketitle


\section{Introduction}

Quantum error mitigation (QEM) \cite{cai2022quantum,
qin2022overviewQEM,endo2021hybrid,temme2017ZNEPECorig,
quek2024exponentially,takagi2022fundamental,li2017efficient,
endo2018practical,strikis2021learning,czarnik2021error,
Koczor2021,Huggins2021,giurgica2020digital,
cai2021symmetry,mari2021extending,huo2022DualState,lowe2021unified,nation2021scalable,
liao2024MLqem,bravyi2021mitigating,kim2023scalable,berg2022probabilistic,
kandala2019error,song2019quantum,arute2020hartree,urbanek2021mitigating,
zhang2020error,sagastizabal2019experimental,he2020zero,
TEMqem,kim2020neural,cantori2024synergy,araki2025spacetimeVD,
bako2025exponentialDualState} has emerged as a practical solution for removing noise bias from expectation values in noisy quantum computer experiments 
\cite{Huggins2021,kim2023scalable,kandala2019error,song2019quantum,arute2020hartree,urbanek2021mitigating,zhang2020error,sagastizabal2019experimental,Quantinuum2025QuantumMagnetismPEA,kim2023evidence,kim2024IBMstabilizeNoisePEC}. 
Unlike quantum error correction (QEC), which requires substantial hardware overhead, specifically, an increase in the number of qubits, most QEM methods require no hardware overhead. Instead, the cost is in runtime, i.e., QEM incurs a sampling overhead that manifests itself as an increase in the number of shots required for a given target accuracy.

However, QEM is not a solution to the scalability problem. As the circuit volume increases linearly with a given hardware platform, the sampling overhead grows exponentially. Thus, with current noise levels, QEM is applicable primarily to circuits with up to a hundred qubits in superconducting circuits. 
Consequently, QEM is often regarded as an intermediate solution until fault-tolerant quantum computing becomes practical. Yet, QEC, even when operating at scale, will not be strictly perfect. It will still struggle with leakage errors, correlated errors, and coherent errors, all of which are challenging for QEC codes to correct. 
Additionally, imperfect ancilla preparation and measurement errors significantly complicate the decoding process.

It has been suggested that hybrid approaches combining QEC and QEM can avoid the problems inherent to each approach \cite{suzuki2022quantum,2021errorQEMforTgatesQEC,lostaglio2021error,
Qedma2025ImportanceConcept,wahl2023ZNEforQEC}. 
The key idea is that QEC will perform most of the heavy lifting by removing local, uncorrelated errors such as decoherence (T2) and amplitude damping (T1), which lead to the exponential overhead when QEM is used without QEC. QEM, on the other hand, will address all the correlated errors and coherent errors that are difficult or impossible to correct with the selected QEC code.

Our approach for QEC-QEM integration is to construct a QEM method that is applicable to any dynamic circuit. As such, it will also mitigate expectation values obtained from error-corrected circuits, since QEC codes are instances of dynamic circuits. Furthermore, dynamic circuits may offer substantial advantages even in the absence of error correction. Examples include the semi-classical Fourier transform \cite{baumer2024scQFT}, gate teleportation \cite{baumer2024LongRange}, entangled states preparation \cite{baumer2024LongRange,Wallraff2024MCMprepFanOut,
Circa2024MCMpred,Buhrman2024MCMprep,yu2024MCMprep,Quantinuum22023AdaptiveAdvantage,Cirac2021adaptiveCircPrep}, and variational algorithms \cite{deshpande2024dynamicNOplateau,xie202mid-circBarren}. 
Therefore, a mitigation method compatible with dynamic circuits is valuable also for the current, pre-fault-tolerant stage of quantum computing.

A key desired feature of QEM protocols is drift resilience. During the course of the experiment, the noise parameters typically vary over time, e.g. changes in the decoherence time that occur due to the activation of two-level system defects in superconducting circuits. Noise drifts can also occur due to temporal variations of coherent errors, such as crosstalk and miscalibration, that, in turn, get converted to time-dependent incoherent errors by twirling techniques \cite{PhysRevX.11.041039,santos2024pseudo,wallman2016noise}. 
Noise drifts are often slow relative to the time scale of a single shot but faster than the duration of the whole experiment. QEM methods can increase the overall runtime by an order of magnitude or more, thereby increasing the likelihood of noise drifts. Consequently, such drifts can have a substantial impact on the final outcomes. Finally, the target accuracy is also important. It is possible that on a given platform, the chosen target accuracy is sufficiently low such that noise drifts can be ignored, but as the platform evolves and the target accuracy becomes more stringent, these drifts may become an obstacle.

Among existing QEM schemes, only a few have the potential to be truly resilient to noise drifts. Methods relying on noise characterization, such as PEC, PEA, Clifford regression, and machine learning, are inherently sensitive to temporal noise drifts. 
Methods based on agnostic, i.e., characterization-free, amplification of noise, such as pulse-stretching zero-noise extrapolation (ZNE)\cite{temme2017ZNEPECorig,kim2023scalable}, digital ZNE \cite{he2020zero,majumdar2023bestPracticeDZNE,PhysRevA.105.042406}, NOX \cite{ferracin2024NOX}, and the adaptive KIK \cite{npjqiKIK} method, can be executed in a drift-resilient manner, as described in ref.~\cite{npjqiKIK} and in Supplementary Note \ref{SupplementaryNote1} of the present paper. 
Pulse stretching requires careful control and calibration procedures. 
Furthermore, it is also not fully compatible with twirling techniques such as randomized compiling and pseudo-twirling, since the stretching mis-scales coherent errors. 
Digital ZNE exhibits a strong error bias at any mitigation order when the noise does not commute with the ideal unitary, which is typically the case; For experimental demonstrations, see Supplementary Note \ref{SupplementaryNote1} of the present paper, and Fig.~3 of ref. \cite{npjqiKIK}. 
NOX is a first-order mitigation theory, and therefore, it is limited to weak noise scenarios.

The Adaptive KIK, on the other hand, is drift-resilient and has convergence assurances when the noise is not very strong. Therefore, it stands as a strong and possibly the only candidate for drift-resilient integration of QEM with QEC. 
The term “adaptive” denotes an efficient post-processing technique that outperforms the standard ZNE post-processing. 
“KIK” refers to the circuit construction and execution, as well as the manipulation of the noise. 
Adaptive post-processing improvements that reduce the sampling overhead by orders of magnitude have been studied in \cite{uzdin2026orders}.

On top of noise drifts, there is another important challenge that arises in some QEM and QEC integration schemes. One suggestion for combining them is to apply QEM to the logical qubits \cite{Qedma2025ImportanceConcept}. 
However, characterization-based error mitigation methods require that 
1) the errors be sufficiently pronounced to be learned within a reasonable timeframe with adequate accuracy, and 
2) that the noise structure be simple enough to be described using a small number of parameters. 
In the NISQ era, two-qubit gate errors are large enough to be learned quickly. This is no longer true for logical two-qubit gates, where errors are expected to be minuscule. 
Such tiny error rates require time-consuming high-accuracy noise characterizations. To avoid this problem, it is possible to apply error mitigation to deeper circuit layers, where the errors are more pronounced. 
However, this poses a challenge, since deeper layers typically do not correspond to Clifford circuits, unlike a single layer of two-qubit gates, which are chosen to be Clifford gates in characterization-based mitigation. When a layer contains only Clifford multi-qubit gates, Pauli-twirling can be used to make the noise more sparse and reduce the characterization time. 
While pseudo twirling \cite{santos2024pseudo} can be used to simplify the noise in non-Clifford layers, it is unclear whether it enables successful characterization-based error mitigation. 
It appears then that QEC is incompatible with the two requirements of characterization-based QEM mentioned above.

Although the KIK protocol, as presented in ref.~\cite{npjqiKIK} is not affected by characterization issues described above, there are two intrinsic hurdles in combining the KIK method with QEC. 
First, it is incompatible with mid-circuit measurements, which are essential for executing syndrome measurements in QEC protocols. 
Second, it exhibits a small bias due to high-order Magnus expansion corrections. 
Such corrections can become significant when the noise is strong or when high accuracy is required. The present work shows how a layer-based application of KIK overcomes these two hurdles simultaneously and opens a path towards drift-resilient QEC-QEM protocols.

When the ``Layered KIK'' (LKIK) protocol is combined with the agnostic amplification readout mitigation presented in ref. \cite{2024E2Eparity}, an end-to-end mitigation scheme emerges, in which preparation, gate, measurement and mid-circuit measurement errors are all treated on the same footing, and in a drift-resilient manner. We point out that LKIK can be combined with characterization-based methods such as PEC, PEA, machine learning, and others. 
Although not drift-resilient and limited to a sparse noise model, these methods can reduce the bulk of the noise, while LKIK can address noise drifts and errors that fall outside the assumed sparse noise model. 
Therefore, LKIK should be evaluated not only as a standalone method, but also as a potential upgrade to existing approaches.

\section{Results}

\subsection{Noise amplification vs noise characterization}

QEM methods can be broadly divided into those that require noise characterization and those that are agnostic to the noise structure. The first category includes probabilistic error cancellation \cite{temme2017ZNEPECorig,mari2021extending,PECtimeDrift}, probabilistic error amplification \cite{kim2023evidence,Quantinuum2025QuantumMagnetismPEA}, Clifford regression \cite{strikis2021learning,czarnik2021error} and machine learning \cite{liao2024MLqem,muqeet2024QLEAR,kim2020neural}. 
The second category includes zero noise extrapolation (ZNE) \cite{temme2017ZNEPECorig}, 
purification \cite{Koczor2021,Huggins2021,araki2025spacetimeVD}, and the Adaptive KIK \cite{npjqiKIK} method. 
For a noise-agnostic readout mitigation, see ref.~\cite{2024E2Eparity}.

Noise characterization approaches can offer lower sampling overhead when noise parameters are time-independent; however, they introduce intrinsic sensitivity to temporal noise drifts. The noise parameters typically vary over the course of an experiment. 
For example, decoherence rates may vary throughout the experiment due to temperature variation, stray external electromagnetic fields, two-level system defects in superconducting circuits, and other factors. 
Since error mitigation incurs a sampling overhead, the experiment runtimes may reach dozens of hours or more during which drift effects may become substantial \cite{kim2023evidence,kim2024IBMstabilizeNoisePEC}.

When the noise levels are stronger or when greater target accuracy is required, precise noise characterization becomes more critical. However, improving the characterization accuracy increases the characterization time, which in turn raises the likelihood of noise drifts. 
Consequently, the characterization may no longer reflect the actual noise profile. 
Although in ref.~\cite{PECtimeDrift} predictions about the expected change in the noise parameters were used for reducing the impact of the noise drift on PEC, we argue that this problem persists and worsens as the circuit size and (or) target accuracy grow.

\subsection{Amplification}

Characterization-free mitigation can be implemented either through purification methods \cite{Koczor2021,Huggins2021,araki2025spacetimeVD} or via noise amplification. 
In this work, we focus on noise amplification, since purification involves hardware overhead and is restricted to limited noise models. 
Alternatives involving approximate backward evolution, which require no hardware overhead, were studied in refs.~\cite{huo2022DualState,bako2025exponentialDualState}.

By controllably changing the level of the noise without altering the functionality of the ideal circuit, it is possible to design a set of noisy circuits that serve as a basis for the post-processing construction of the inverse noise channel. 
To state this more formally, we use the Liouville space notation where the $n\times n$ density matrix representing the quantum state of the quantum computer is vectorized into a density vector $\ket{\rho}_{n^{2}\times1}$.
Any linear map acting on $\rho$, e.g. the time evolution operator $K$, can be written as an operator acting from the left: $\ket{\rho(t)}=K\ket{\rho(0)}$. Writing $K$ in terms of the ideal unitary $U$ (in Liouville space) and the noise $N$, we have $K = UN$. 
Amplification-based methods aim to generate a set
\begin{equation}
\{Uf_{j}(N)\}_{j=0}^{M},
\end{equation}
where the functions $f_{j}$ are known precisely, but $N$ is unknown. 
For example, a common choice is $K\in\{UN,UN^{3},UN^{5}\}$. 
Next, a linear combination is constructed such that: 
\begin{equation}
K_{mit}^{(M)}=\sum_{j=0}^{M}a_{i}^{(M)}Uf_{j}(N)\simeq U.
\end{equation}
This is equivalent to $N^{-1}\sum_{j=0}^{M}a_{j}^{(M)}f_{j}(N)\simeq N^{-1}$ which, in the case of odd-powers amplification, reads 
\begin{equation}
\sum_{j=0}^{M}a_{j}^{(M)}N^{2j}\simeq N^{-1}.\label{eq: N-1 expansion}
\end{equation}
Since the coefficients $a_{j}$ have mixed signs, the linear combination must be carried out in post-processing. 
Furthermore, the unavoidable presence of negative coefficients is directly associated with the sampling overhead of the method. 
Equation~\eqref{eq: N-1 expansion} shows that noise mitigation via amplification reduces to a polynomial approximation of the function $g(x)=x^{-1}$. 
When expressed in terms of eigenvalues, this becomes essentially a one-dimensional calculus problem.

\subsection{Taylor and adaptive coefficients}

The simplest polynomial approximation is the Taylor series expansion around $N=I$, where $I$ denotes the identity channel.
By setting $\epsilon=(N^{2}-I)$ and Taylor expanding the expression $N^{-1}=\sqrt{1/(I+\epsilon)}$ 
one can derive an approximation of $N^{-1}$ in terms of $\epsilon$ powers, which can be further expanded and regrouped in powers of $N^{2j}$. 
We refer to the coefficients obtained by doing so as the Taylor coefficients $a_{j,Tay}^{(M)}$
\begin{equation}
a_{j,Tay}^{(M)}=\frac{(-1)^{j}(2M+1)!!}{2^{M}(2j+1)j!(M-j)!}.\label{eq: Taylor coef}
\end{equation}
These coefficients are equivalent to those obtained via Richardson extrapolation (see the discussion in ref.~\cite{npjqiKIK}). 
Notably, the approach described above is, in general, different from the ZNE approach. 
The Taylor coefficients are designed for weak noise, as they best approximate the $N^{-1}$ operator at $\epsilon=0$ (zero noise). 
However, if prior knowledge on the eigenvalues interval is available, more efficient techniques can be used. 
For example, in ref.~\cite{npjqiKIK} the adaptive KIK approach minimizes the $L^2$ norm $\intop_{g(\mu)}^{1}|\sum_{j=0}^Ma_{j}^{(M)}(g)\lambda^{j}-\lambda^{-1/2}|^{2}d\lambda$, where $g$ is estimated from a simple echo measurement. 
The adaptive coefficients $a_{j}^{(M)}(g)$ yield stronger noise mitigation with less shots compared to the Taylor coefficients.
Equivalently, the same mitigation strength can be achieved with a lower mitigation order, i.e., with fewer shots and circuits. 
See \cite{uzdin2026orders} for a simple and effective alternative using virtual noise scaling.

\subsection{Noise-agnostic amplification methods: digital ZNE, adaptive KIK}

Inversion methods based on amplification may employ different techniques to achieve the desired amplification. For example, the original ZNE protocol uses pulse stretching \cite{temme2017ZNEPECorig,kim2023scalable}. 
Subsequently, digital amplification schemes based on gate replication (also known as local folding or gate insertion) or on full circuit amplification have been suggested \cite{majumdar2023bestPracticeDZNE,he2020zero,PhysRevA.105.042406}. 
However, it has been shown that digital noise amplification is correct only in the uncommon cases where the noise happens to commute with the ideal gate, as in the global depolarizing noise channel.
These digital amplification methods rely on the fact that in the absence of errors, CNOT gates or CPhase gates are their own inverse and therefore $U_{gate}^{3}=U_{gate}$. 
Based on this, these methods assume that $K^{3}\simeq U_{gate}N_{gate}^{3}$ where $K\simeq U_{gate}N_{gate}$ is the original noisy gate.
According to this assumption, the original noise channel was amplified by a noise amplification factor of three. 
However, a simple calculation shows that $K^{3}=U_{gate}N_{gate}\tilde{N}_{gate}N_{gate}$
where $\tilde{N}_{gate}=U_{gate}N_{gate}U_{gate}$. 
This generates an error bias at leading order in the noise. 
See ref.~\cite{npjqiKIK} for an analytical and experimental verification of the failure of gate insertion amplification. 
An additional experimental verification is given in Supplementary Note \ref{SupplementaryNote1} of the present paper.

The problem of gate insertion can be resolved by using a pulse-based inverse protocol that takes time ordering into account \cite{npjqiKIK}. 
The pulse inverse protocol is obtained by reversing the schedule of the (effective) interaction Hamiltonian and inverting its sign. 
Crucially, this is done even for gates that are equal to their own inverse, such as CNOT and CPhase. 
We denote the pulse inverse of a circuit $K$ by $K_I$. 
The frequent appearance of $K_IK$ terms in this framework led to the name ``KIK''. 
The adaptive KIK method is also valid for the Mølmer-Sørensen gate in trapped ions. 
In Supplementary Note \ref{SupplementaryNote1} we provide a trapped ions experimental demonstration.

In ref.~\cite{npjqiKIK}, the pulse-inverse evolution is defined for the circuit as a whole (``global folding'') rather than for individual gates or layers. 
This is the natural starting point when using the Magnus expansion formalism. 
In this paper, we study the advantages and the analytical structure of layer-based KIK (LKIK). 
We refer to the version described in ref.~\cite{npjqiKIK} as ``Global KIK'' (GKIK). 
In fact, the GKIK can be considered as special case of LKIK, where a single layer contains the whole circuit. 
In the context of digital ZNE, a comparison between local and global folding noise amplification has been carried out in ref.~\cite{schultz2022impact}. 
While the findings favored global folding, it should be noted that the digital ZNE amplification scheme is incorrect if the noise does not commute with the ideal unitary operation. Moreover, the study did not consider twirling protocol for suppressing coherent errors. In particular, pseudo-twirling becomes more efficient as the number of layers increases \cite{santos2024pseudo}.

\subsubsection{Features of GKIK QEM}

The adaptive KIK seamlessly integrates with methods for treating coherent errors such as Pauli twirling, randomized compiling (RC) \cite{wallman2016noise,PhysRevX.11.041039,Emerson2020LearnTwirlExpIBM,knill2004PauliTwirl,RCqutrits}, and pseudo-twirling (PST) \cite{santos2024pseudo,chen2022SimPSTwrong,layden2023MarkovMontePST,Quantinuum2025QuantumMagnetismPEA}. 
While RC works also for single-qubit non-Clifford gates, PST also handles multi-qubit non-Clifford gates and consequently facilitates shorter, less noisy circuits. 
Using PST, it has been shown that adaptive KIK can be used for high-accuracy gate calibrations. 
In particular, the calibration of a sequence of 81 $ZX(\pi/2)$ gates (equivalent to 81 CNOTs up to single-qubit gates) was demonstrated.

As previously noted, one of the major advantages of the GKIK protocol is that it is drift-resilient. 
It allows noise parameters to vary over time without any degradation in performance. 
The only assumption needed is that these noise parameters remain stable over the duration of several shots (approximately twenty, for mitigation order three or lower). 
We distinguish the noise drift time dependence from a different type of time dependence, arising from the fact that different gates may consistently experience different types of noise. 
This faster, intra-shot time-dependence is automatically addressed by the GKIK protocol. 
An experiment demonstrating the GKIK drift resilience is described in Supplementary Note \ref{SupplementaryNote1}. 
As explained there, drift resilience does not automatically follow from the noise agnostic amplification and the absence of noise characterization. 
Yet, by applying a specific execution order for the various noise amplification circuits, temporal noise drift effects are fully eliminated.

Nevertheless, the GKIK method has its own flaws. 
First, it requires inverting the sign of the effective Hamiltonian, which in some platforms is not trivial to implement. In the cross-resonance gate and the MS gate, virtual Z gates can easily be used to achieve this inversion (see Supplementary Note \ref{SupplementaryNote2}), and no pulse-level programming is required. 
However, for fixed or tunable coupler phase gates \cite{krantz2019QuantEng}, it is less straightforward how to reliably implement the pulse inverse. 
Interestingly, it has been suggested that cross-resonance gates can be implemented in tunable-qubit setups \cite{xu2023parasiticCR}. 
Hence, the GKIK method might be relevant for tunable coupler platforms as well. 
As a method for integration of KIK with QEC, the GKIK protocol also has the two limitations which were mentioned earlier: 
i) The global folding used in GKIK is incompatible with mid-circuit measurements. 
ii) The KIK formulation in ref.~\cite{npjqiKIK} is based on neglecting high-order Magnus noise terms. 
This can pose a problem in strong noise scenarios or, alternatively, when high accuracy is required, as demonstrated in section ``Problems with the Global KIK mitigation''.

In this work, we show how these two limitations can be simultaneously resolved by adopting a layer-based KIK (LKIK) scheme. While the LKIK circuits are different from those used in the GKIK protocol, they incur the same sampling overheads and exploit the same post-processing (excluding the adaptive post-processing used in ref.~\cite{npjqiKIK}). 
Therefore, the improvement comes without any additional cost of sampling or experimental complexity. To the best of our knowledge, this work presents the first drift-resilient and bias-free gate error mitigation in dynamic circuits.

In a companion work focusing primarily on mid-circuit measurement error mitigation \cite{2024E2Eparity}, we experimentally demonstrate the effectiveness of LKIK in mitigating errors in dynamic quantum circuits.

\subsection{A recap of the GKIK method and its limitations}

\subsubsection{Recap of the GKIK method}

Given an observable of interest $A$, initial state $\rho_{0}$, and a circuit to be mitigated with a noisy evolution operator $K$, the noisy expectation value is given by $\braOket AK{\rho_{0}}$ Liouville space notation \cite{gyamfi2020fundamentals}. 
The GKIK method requires the execution of circuits of the form $K(K_{I}K)^{j}$ and the measurement of their associated expectation values
\begin{equation}
A_{j}=\braOket A{K(K_{I}K)^{j}}{\rho_{0}}.
\end{equation}
The $M-$th order mitigated expectation value is obtained via 
\begin{equation}
A_{mit}^{(M)}=\sum_{j=0}^{M}a_{j}^{(M)}[g(\mu)]A_{j},\label{eq: GAK sum}
\end{equation}
where the adaptive coefficients $a_{j}^{(M)}(g)$ are known analytically and depend on the magnitude of the circuit echo $\mu$
\begin{equation}
\mu=\braOket{\rho_{0}}{K_{I}K}{\rho_{0}},
\end{equation}
$\mu$ is evaluated by executing an additional circuit of the form KIK. The non-adaptive version of the global KIK method is obtained by setting $g = 1$. 
It yields the Taylor coefficients and does not require the execution of the echo circuit. Substantially stronger mitigation with fewer circuits and lower shot overhead is achieved by setting $g = \mu^2$. 
See \cite{uzdin2026orders} for an alternative that does not require an echo measurement.

\subsubsection{Noise drift resilience}

To achieve resilience to noise drift, the execution order of the circuits is structured as follows: the total number of shots is divided into $N$ sets, where $N$ is chosen such that the drift in the noise parameters is negligible during the execution of each set.
In each set, all amplification circuits $K(KIK)^{j}$ are executed and Eq.~\eqref{eq: GAK sum} is evaluated. 
Each set may experience a different bias (without mitigation) due to the temporal drift of the noise parameters. 
This bias leads to different $A_j$ values in each set, but the sum Eq.~\eqref{eq: GAK sum} removes this bias from each set. 
Since the number of shots in each set is small, the variance of \eqref{eq: GAK sum} in each set is large. 
After averaging over all sets, the variance is reduced. 
This is experimentally demonstrated in Supplementary Note \ref{SupplementaryNote1} along with a more rigorous explanation.

\subsection{Problems with the global KIK mitigation}

\subsubsection{Incompatibility of GKIK with mid-circuit measurement}

The neglect of high-order Magnus term discussed in the next subsection implies that GKIK cannot treat arbitrarily strong noise. 
Elements such as projective mid-circuit measurement (MCM) that completely decohere the state are equivalent to infinitely strong noise and therefore cannot be described within the global folding description of the GKIK framework. 
More importantly, the projection effect is not something we wish to mitigate. 
It is part of the ideal dynamic circuit, and we wish to keep it intact during the mitigation. 
Thus, in the GKIK construction, it is unclear how to construct the pulse-inverse circuit while retaining the original mid-circuit functionality. 
For example, for the circuit 
$K_{2}\mathbb{M}K_{1}$
where 
$\mathbb{M}$
is a projection operator that describes a measurement, choosing 
$K_{1}^{I}\mathbb{M}K_{2}^{I}$
as the pulse inverse leads to incorrect noise amplification.

\subsubsection{Bias originating from high-order Magnus noise terms}

The GKIK does not assume a specific noise structure, but it does rely on two assumptions regarding the character of the noise and its magnitude. 
The first assumption is Markovianity. 
However, if the noise exhibits pronounced non-Markovian features, it is possible to suppress them by applying twirling schemes such as RC \cite{wallman2016noise,PhysRevX.11.041039,Emerson2020LearnTwirlExpIBM,knill2004PauliTwirl,RCqutrits,cai2019SmallerTwirl} or PST \cite{santos2024pseudo,chen2022SimPSTwrong,layden2023MarkovMontePST,Quantinuum2025QuantumMagnetismPEA}. 
The second assumption is that high-order Magnus noise terms are negligible. 
In practice, their effect can be challenging to observe. 
They manifest either when the noise is very strong, in which case the sampling cost is sometimes already unrealistic, or when very high accuracy is needed, in which case the sampling cost required to achieve this accuracy is likewise unrealistic.

Expressing $K$ in terms of the Magnus expansion of the interaction picture in Liouville space \cite{npjqiKIK}, we get
\begin{equation}
K=Ue^{\sum_{i=1}^{\infty}\Omega_{i}^{G}},
\end{equation}
where
\begin{align}
\Omega_{1}^{G} & =\int_{0}^{\tau}dt\mc L_{int}(t)dt,\\
\Omega_{2}^{G} & =\frac{1}{2}\int_{0}^{\tau}dt'\int_{0}^{t'}dt[\mc L_{int}(t'),\mc L_{int}(t)],
\end{align}
and 
\begin{equation}
\mc L_{int}(t)=U^{\dagger}(t)\mc L(t)U(t).
\end{equation}

\begin{figure}[h!]
\includegraphics[width=\linewidth]{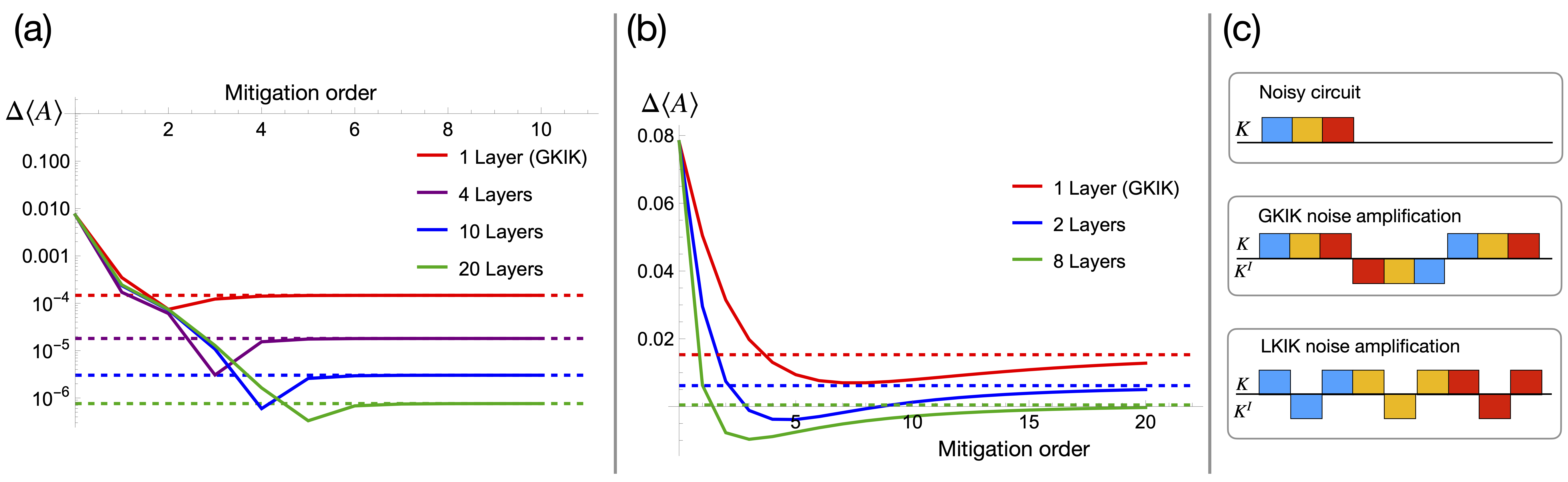}
\caption{\label{fig:A-four-qubit-simulation}A four-qubit simulation demonstrating the advantage of using the layered-based KIK (LKIK) amplification over the global KIK amplification (GKIK - single layer).
In (a) and (b) the y axis is the difference $\Delta \langle A\rangle$ between the mitigated expectation value (see main text) and the ideal value, and the $x$ axis is the mitigation order $M$ with $M=0$ indicating no mitigation. 
The ideal expectation value is $\langle A\rangle \simeq 0.025$. 
In (a) the noise strength parameter is $\xi = 0.02$ and in (b) it is $\xi=0.2$. 
Figure (a) shows that LKIK is essential for achieving high accuracy even when the noise is weak, while Fig.~(b) shows that LKIK is important when the noise is strong even when the requested target accuracy is modest. 
The dashed lines show the prediction of the Layered KIK formula \eqref{eq: Layered KIK formula}. 
(c) An illustration of a three-layer circuit (top) being noise amplified with GKIK (middle) and LKIK (bottom). 
The amplification factor is three for both cases. 
Squares below the black horizontal line represent pulse-inverse operation.}
\end{figure}

The superscript ``G'' denotes ``Global'' referring to the fact that these Magnus terms include the whole circuit and not just parts of it. 
In ref.~\cite{npjqiKIK} it has been shown that $K_{I}K=e^{2\Omega_{1}^{G}}+O(\Omega_{3}^{G})$, and as a result, the Global KIK formula is obtained
\begin{equation}
\label{eq: GKIK formula}
K_{mit}^{(\infty)}=K\frac{1}{\sqrt{K_{I}K}}=U+O(\Omega_{2}).
\end{equation}
As the mitigation order $M$ increases, $K_{mit}^{(M)}=\sum_{j=0}^{M}a_{j}^{(M)}K(K_{I}K)^{j}$
approaches $K\frac{1}{\sqrt{K_{I}K}}$. 
The red curves in Fig.~\ref{fig:A-four-qubit-simulation}a, b show the deviation of the mitigated expectation value $A$ from the ideal value, 
$\Delta\left\langle A\right\rangle =\left\langle A\right\rangle _{mit}-\left\langle A\right\rangle _{ideal}$, 
as a function of the mitigation order, where order zero corresponds to no mitigation. 
The observable is 
$A=\ketbra{\psi_{0}}{\psi_{0}}$, 
which captures the population of the initial state 
$\ket{\psi_{0}}=\ket{0000}$ at the end of the evolution. 
The system comprises a chain of four qubits with $\sigma_{x}\otimes\sigma_{x}$ interaction between nearest neighbors, such that
$H=\sigma_{x}^{(1)}\otimes\sigma_{x}^{(2)}+\sigma_{x}^{(2)}\otimes\sigma_{x}^{(3)}+\sigma_{x}^{(3)}\otimes\sigma_{x}^{(4)}$. 
The evolution time is $\tau=1$, and each qubit is subjected to local decoherence, where the decay coefficient in front of the Lindbladians is $\xi=0.02$ for Fig.~\ref{fig:A-four-qubit-simulation}a and $\xi=0.2$ for Fig.~\ref{fig:A-four-qubit-simulation}b. 
Taylor coefficients are used in both subplots. 
$L = 1$ corresponds to GKIK. 
The dashed line for $L = 1$ shows that the prediction of Eq.~\eqref{eq: GKIK formula} describes well the performance saturation for larger mitigation orders.

\subsection{Layered-based KIK noise amplification}

In what follows, we introduce the notion of KIK amplification in layers and show that the errors associated with $\Omega_2$ are suppressed as the number of layers $L$ increases. 
In the limit of thin layers, we find that the $\Omega_2$ contribution scales as $1/L^2$. 
We begin by dividing the circuit into non-overlapping time layers. 
The set of layers encompasses the full circuit. 
There is no restriction on the length of individual layers and they can have different lengths. 
A layer may contain sequences of gates, or alternatively, the length of a layer can be a fraction of the duration of a single pulse used to create a gate.

\subsubsection{Introducing the Layered KIK}

Let $a, b$ and $c$ be three layers that compose a larger circuit.
For the noisy circuit, it holds that $K_{tot}=K_{c}K_{b}K_{a}$, where we neglect non-Markovian noise that carries over from one layer to another. Correspondingly, the global pulse inverse of the circuit is $K_{tot}^{I}=K_{a}^{I}K_{b}^{I}K_{c}^{I}$ and the GKIK mitigated evolution has the form
\begin{equation}
K_{mit,GKIK}^{(M)}=\sum_{j=0}^{M}a_{j}^{(M)}K_{c}K_{b}K_{a}(K_{a}^{I}K_{b}^{I}K_{c}^{I}K_{c}K_{b}K_{a})^{j}.
\end{equation}
In this paper, we suggest a layer-based amplification where the evolution operator of $l-$th layer, $K_l$, is replaced by $K_{l}(K_{l}^{I}K_{l})^{j}$, yielding a noise amplification factor of $(2j + 1)$ for each layer. 
In the three-level example, the Layered KIK (LKIK) mitigated evolution operator reads
\begin{align}
K_{mit,LKIK}^{(M)}&=\sum_{j=0}^{M}a_{j}K_{c}(K_{c}^{I}K_{c})^{j}K_{b}(K_{b}^{I}K_{b})^{j}K_{a}(K_{a}^{I}K_{a})^{j} 
\nonumber\\
&=\sum_{j=0}^{M}a_{j}\Pi_{l=a}^{c}K_{l}(K_{l}^{I}K_{l})^{j}.
\end{align}
The two schemes are illustrated in Fig.~\ref{fig:A-four-qubit-simulation}c. 
Note that the layers are not mitigated individually, but only individually amplified. 
To show the relation between the GKIK and LKIK and also how the local amplification of individual layers is related to the full Magnus expansion of all the layers combined, we write
\[
K(K^{I}K)^{j}=Ue^{(2j+1)\Omega_{1}^{G}+\Omega_{2}^{G}+...}.
\]

By considering the Magnus expansion of individual layers, in Supplementary Note~\ref{SupplementaryNote4}, we show that in the absence of mid-circuit measurements, it holds that $\Omega_{1}^{LKIK}=\Omega_{1}^{GKIK}$. 
Therefore, under the approximation that neglects $\Omega_{2}$ and higher order terms, and in the absence of mid-circuit measurement, layer-based noise-amplification and global noise-amplification are the same. 
This result also implies that in the absence of mid-circuit measurements, the same Taylor coefficients and adaptive coefficients can be used in both schemes.Before examining the second-order Magnus terms, it is instructive to study how the Taylor coefficients mitigate various orders of noise arising from different layers. 
This analysis will facilitate the inclusion of mid-circuit measurements or other potential elements that should not be mitigated. 
As shown in Supplementary Note~\ref{SupplementaryNote5}, the Taylor coefficients $a_{j}^{(M)}$ have the following properties

\begin{align}
\sum_{j=0}^{M}a_{j}^{(M)}(2j+1)^{m} & =0, \text{\ \ for 1\ensuremath{\le m\le M}},\\
\sum_{j=0}^{M}a_{j}^{(M)} & =1.
\end{align}

As a result, for an analytic function $F(x)$ at $x=0$ it holds that 
\begin{align}
\sum_{j=0}^{M}a_{j}^{(M)}F[(2j+1)x] & =\sum_{k=0}^{M}a_{j}^{(M)}\sum_{m=0}^{\infty}\frac{F^{(m)}(0)}{m!}(2j+1)^{m}x^{m}\nonumber \\
& =\sum_{m=0}^{\infty}\frac{F^{(m)}(0)}{m!}x^{m}\sum_{j=0}^{M}a_{j}^{(M)}(2j+1)^{m}\nonumber \\
& =F(0)+O(x^{M+1}).
\end{align}
When this is applied to $F(\Omega_{1})=Ue^{\Omega_{1}}$, we obtain the Taylor mitigation for a single layer (GKIK). Next, we consider the case:
\begin{align}
\sum_{j=0}^{M}a_{k}^{(M)}F_{2}[(2j+1)x_{2}]F_{1}[(2j+1)x_{1}] & =\sum_{k=0}^{M}a_{k}^{(M)}\sum_{m_{2}=0}^{\infty}\frac{F_{2}^{(m_{2})}(0)}{m_{2}!}(2j+1)^{m_{2}}x_{2}^{m_{2}}\sum_{m_{1}=0}^{\infty}\frac{F_{1}^{(m_{1})}(0)}{m_{1}!}(2j+1)^{m_{1}}x_{1}^{m_{1}}\nonumber \\
 & =\sum_{m_{1},m_{2}=0}^{\infty}\frac{F_{2}^{(m_{2})}(0)}{m_{2}!}\frac{F_{1}^{(m_{1})}(0)}{m_{1}!}x_{2}^{m_{2}}x_{1}^{m_{1}}\sum_{j=0}^{M}a_{k}^{(M)}(2j+1)^{m_{1}+m_{2}}\nonumber \\
 & =F_{2}(0)F_{1}(0)+O(x_{2}^{m_{2}}x_{1}^{m_{1}})|_{m_{1}+m_{2}>M}.
\end{align}
To apply this to the LKIK mitigation of two layers we replace the functions by operators:
\begin{align}
F_{l}[(2j+1)x_{l}]\to F_{l}[(2j+1)\Omega_{1,l}]&=U_{l}e^{(2j+1)\Omega_{1,l}}
\nonumber\\ 
&=K_{l}(K_{l}^{I}K_{l})^{j}+O[\Omega_{2,l}].
\end{align}
Time ordering is respected since ordering was also kept in the multivariable analysis. Consequently, the mitigated LKIK evolution operator satisfies 
\begin{align}
K_{mit,LKIK}^{(M)}&=\sum_{j=0}^{M}a_{j}^{(M)}\Pi_{l=1}^{L}K_{l}(K_{l}^{I}K_{l})^{j}
\nonumber\\
&=\Pi_{l=1}^{L}U_{l}+O(\Pi_{l}\Omega_{1,l}^{m_{j}}|_{\sum m_{j}>M},\Omega_{2,l}).
\end{align}
This result can be generalized to any number of layers. 
We conclude that all elements that scale appropriately are mitigated by the Taylor coefficients. 
For example, the quadratic term arising from $\Omega_1^2$
of a specific layer, and the product of two $\Omega_1$ terms from different layers, both scale the same when the noise is amplified. 
As such, both will be mitigated when the mitigation order is two or higher.

As shown in Fig.~\ref{fig:A-four-qubit-simulation}a, the accuracy improves as the number of layers increases. 
The same behavior is observed at lower target accuracies when the noise is strong (Fig.~\ref{fig:A-four-qubit-simulation}b). 
Both cases illustrate that increasing the number of layers significantly enhances performance while the sampling overhead is the same, since in both cases the same Taylor coefficients are used at each order. The sampling overhead and the mitigation order can be substantially improved by using adaptive coefficients, as described in ref.~\cite{npjqiKIK}. 
To reduce the initial error bias by a factor of 200, for example, order 19 is needed for the Taylor coefficients, and order 9 for the adaptive coefficients. While the Taylor coefficients incurs a sampling overhead of 138,852, the overhead in the adaptive case is 9873. 
We point out that sampling overheads on the order of ten thousand or more are realistic for QEM protocols, which are drift-resilient. On top of that, in bias-free methods, the shots can be collected in parallel on different quantum processors, for example using 50 computers with a sampling overhead of 100 for each. Finally, by applying error correction before the LKIK protocol, the noise mitigation sampling overhead is expected to be smaller, since most of the noise will be removed by the error correction procedure.

\subsubsection{Incorporating mid-circuit measurements}

Next, we add an element $\mc{\mathbb{M}}$ that represents an operation that should not be mitigated. 
For example, a mid-circuit measurement can be represented by a sum of projection operators in Liouville space $\mathbb{M}=\sum_{k}\mc{\mathbb{M}}_{k}$.
A feedforward evolution operator can be written as 
$$
K=\sum_{k}K_{b,k}\mc{\mathbb{M}}_{k}K_{a}.
$$
Here, layer $b$ is conditioned on the outcome of the measurement following layer $a$.
Operationally, when the amplification level is $j$ and the measurement outcome is $k$, a layer $K_{b,k}(K_{b,k}^{I}K_{b}{}_{k})^{j}$ will be executed. 
Thus, the mitigated evolution operator takes the form
\begin{equation}
K_{mit,layers}^{(M)}=\sum_{k}\sum_{j=0}^{M}a_{j}^{(M)}K_{b,k}(K_{b,k}^{I}K_{b,k})^{j}\mc{\mathbb{M}}_{k}K_{a}(K_{a}^{I}K_{a})^{j}.
\end{equation}
Due to linearity, we can treat each mid-circuit measurement (MCM) outcome $k$ separately. 
To include MCM’s in our analysis, we return to the multi-variate point of view and consider cases where some of the functions $F_l$ are not scaled. 
For example, in the case of three layers where only $F_2(x)$ is not scaled, we get
\begin{equation}
\sum_{j=0}^{M}a_{j}^{(M)}F_{3}[(2j+1)x_{3}]F_{2}[x_{2}]F_{1}[(2j+1)x_{1}]=F_{3}(0)F_{2}(x_{2})F_{1}(0)+O(x_{3}^{m_{3}}F_{2}(x_{2})x_{1}^{m_{1}})|_{m_{1}+m_{3}>M}.
\end{equation}
Hence, the scaled functions get mitigated while the unscaled function $F_2$ remains unchanged. 
To include MCM’s, we set 
$$
F_{1(3)}[(2j+1)x]\to K_{1(3)}(K_{1(3)}^{I}K_{1(3)})^{k},F_{2}(x_{2})=\mc{\mathbb{M}}_{k},
$$
and finally obtain 
\begin{equation}
K_{mit,layers}^{(M)}=\sum_{k}\sum_{j=0}^{M}a_{j}K_{k}(K_{k}^{I}K_{k})^{j}\mc{\mathbb{M}}_{k}K_{a}(K_{a}^{I}K_{a})^{j}=\sum_{k}U_{k}\mc{\mathbb{M}}_{k}U_{a}+O(\Omega_{1,3}^{m_{3}}\mc{\mathbb{M}}_{k}\Omega_{1,1}^{m_{1}})|_{m_{1}+m_{3}>M}.
\end{equation}
This shows that LKIK works equally well for dynamic circuits, and it is therefore compatible with quantum error correction codes.
In practice, the measurements are often noisy, which may lead to execution of the wrong gates. 
While standard error readout mitigation techniques designed for terminating measurement do not apply, in refs.~\cite{hashim2025PECreadout,Temme2024dynamicPEC,koh2026readout} it was demonstrated that such errors can be addressed using measurement PEC. 
A drift-resilient alternative was studied in ref.~\cite{2024E2Eparity}. 
This method also mitigates terminating measurement errors and preparation errors. 
An experimental demonstration of LKIK in dynamic circuit requires a proper treatment of mid-circuit measurement errors. 
In a companion paper that deals mostly with mid-circuit measurement error mitigation \cite{2024E2Eparity}, we show an experimental realization of five-qubit dynamic circuit GHZ creation circuit. 
In this experiment, LKIK is used for the gates, and parity-based mitigation for the measurement errors.

\subsubsection{$\Omega_{2}$ Analysis}

\begin{figure}[h!]
\includegraphics[width=10cm]{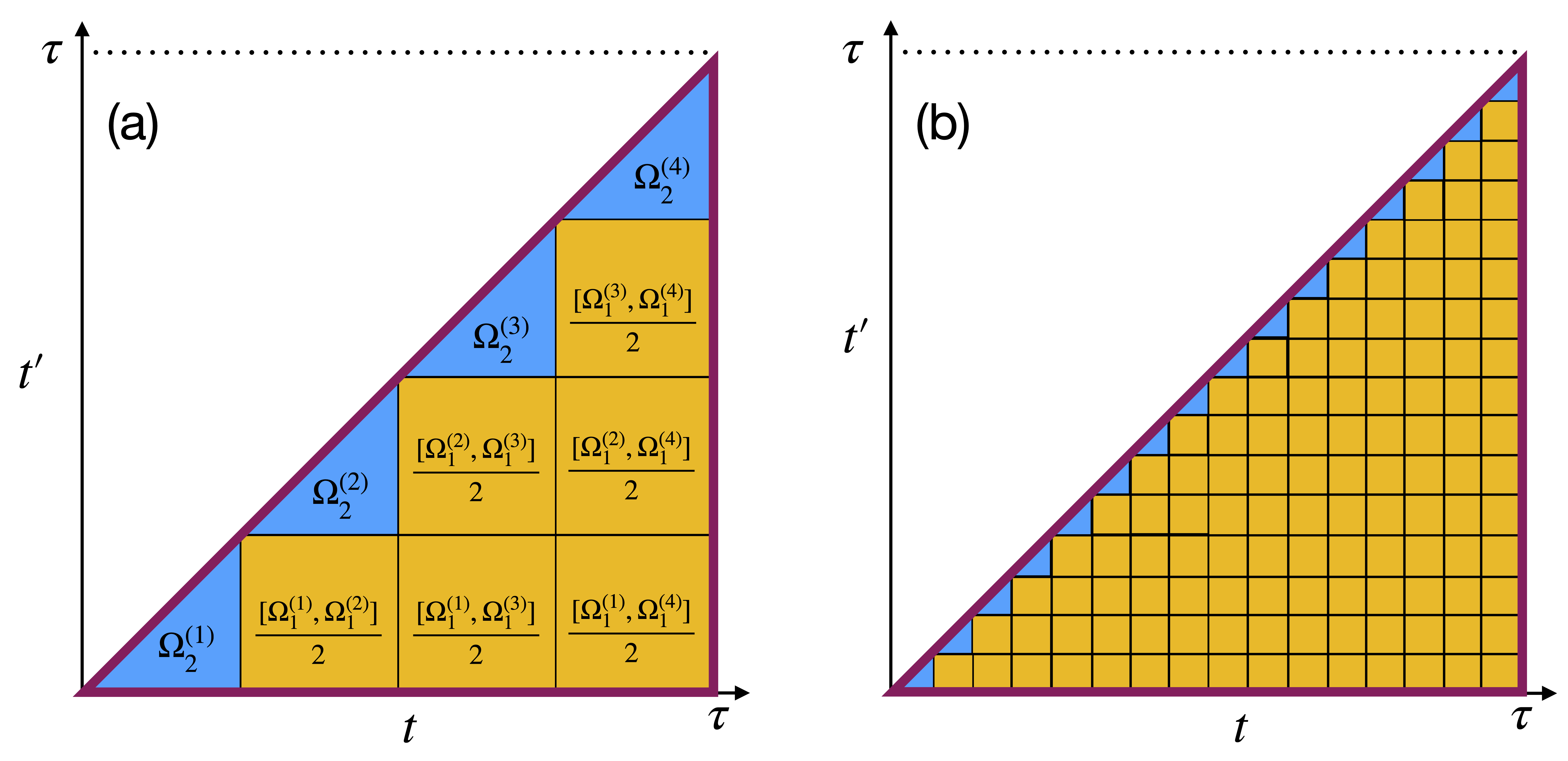}
\caption{\label{fig: illust omega2}Local contributions to the global $\Omega_2$. 
(a, b) The performance of the global KIK introduced in ref.~\cite{npjqiKIK} is limited by the second-order Magnus term of the entire circuit, $\Omega_{2}^{G}$. 
$\Omega_{2}^{G}$ is calculated using a double integral whose integration domain is depicted by the purple triangle in (a). 
$\tau$ is the time duration of the unmitigated circuit. 
The same circuit can be described as a sequence of $L$ consecutive layers ($L = 4$ in (a)). 
As a result, ${\Omega }_{2}^{G}$ can be divided into two different types of contributions: 
i) the blue triangles that arise from the $\Omega_2$ of each layer, and 
ii) the orange squares which originate from the $\Omega_1$ commutator of different layers. 
Crucially, we show that in the Layered KIK protocol, the contribution of the squares is eliminated, leaving only the blue triangles contribution. 
Furthermore, as the layers get thinner (b), the contribution of the blue triangles becomes negligible.
From this argument one can derive an upper bound on the LKIK mitigation error that scales as $1/L$. 
Interestingly, when the layers are sufficiently thin, we find a tighter error bound that scales as $1/L^2$.
As such, LKIK is bias-free in practice, since it is always possible to choose a sufficiently large $L$ that guarantees that the residual mitigation error remains below the target experimental accuracy.}
\end{figure}

Another advantage of LKIK over GKIK concerns the residual bias of the mitigated expectation value relative to the idle expectation value. 
This advantage is unrelated to the MCM advantage described above. 
We start by pointing out that in each amplification level, the total Magnus term $\Omega_2$ has two contributions: one that comes from the $\Omega_2$ of each layer, and the other arising from cross-layer commutators of the $\Omega_1$ terms. 
To see this, we first consider an unamplified circuit. 
Without loss of generality, for three unamplified layers, we have
\begin{align}
U_{c}e^{\tilde{\Omega}_{1,c}+\tilde{\Omega}_{2,c}}U_{b}e^{\tilde{\Omega}_{1,b}+\tilde{\Omega}_{2,b}}U_{a}e^{\Omega_{1,a}+\Omega_{2,a}} & =\nonumber \\
U_{c}e^{\tilde{\Omega}_{1,c}+\tilde{\Omega}_{2,c}}U_{b}U_{a}e^{U_{a}^{\dagger}(\tilde{\Omega}_{1,b}+\tilde{\Omega}_{2,b})U_{a}}e^{\Omega_{1,a}+\Omega_{2,a}} & =\nonumber \\
U_{c}U_{b}U_{a}e^{(U_{b}U_{a})^{\dagger}(\tilde{\Omega}_{1}^{c}+\tilde{\Omega}_{2}^{c})U_{b}U_{a}}e^{U_{a}^{\dagger}(\tilde{\Omega}_{1}^{b}+\tilde{\Omega}_{2}^{b})U_{a}}e^{\Omega_{1,a}+\Omega_{2,a}} & \doteq\nonumber \\
U_{c}U_{b}U_{a}e^{\Omega_{1,c}+\Omega_{2,c}}e^{\Omega_{1,b}+\Omega_{2,b}}e^{\Omega_{1,a}+\Omega_{2,a}}
\end{align}

where the tilde sign refers to quantities that were calculated locally, i.e. the time at the beginning of layer is set to zero. 
Using the Baker–Campbell–Hausdorff (BCH) formula, we get 
\begin{align}
U_{c}U_{b}U_{a}e^{\Omega_{1,c}+\Omega_{2,c}}e^{\Omega_{1,b}+\Omega_{2,b}}e^{\Omega_{1,a}+\Omega_{2,a}} & \cong U_{c}U_{b}U_{a}\nonumber\\
 & \times e^{\Omega_{1}^{tot}+(\Omega_{2,c}+\Omega_{2,b}+\Omega_{2,a})+\frac{1}{2}[\Omega_{1,c},\Omega_{1,b}]+\frac{1}{2}[\Omega_{1,c},\Omega_{1,a}]+\frac{1}{2}[\Omega_{1,b},\Omega_{1,a}]},
\end{align}
where $\cong$ indicates that high-order BCH terms are negelected. 
In Fig.~\ref{fig: illust omega2} we show the local contributions to the global $\Omega_2$. 
The expression 
$(\Omega_{2,c}+\Omega_{2,b}+\Omega_{2,a})+\frac{1}{2}[\Omega_{1,c},\Omega_{1,b}]+\frac{1}{2}[\Omega_{1,c},\Omega_{1,a}]+\frac{1}{2}[\Omega_{1,b},\Omega_{1,a}]$
must be equal to the $\Omega_{2}^{G}$ obtained from a global point of view. 
The triangles represent $\Omega_2$ of each layer, and the squares represent contributions from commutators of $\Omega_1$ across different layers. 
Next, we consider the amplified case and study the $\Omega_2$ terms of the mitigated KIK operators, 
\begin{align}
U{e}^{\alpha {\Omega }_{1}+{\Omega}_{2}+{\Omega }_{3}} & =  \mathop{\sum }\limits_{k=0}^{\infty }
\frac{{(\alpha {\Omega}_{1}+{\Omega}_{2}+{\Omega }_{3})}^{k}}{k!} 
\nonumber\\ 
&
\underset{(M)}{=}\mathop{\sum}\limits_{k=0}^{M}
\frac{\alpha^{k}{\Omega}_{1}^{k}}{k!}+\mathop{\sum}\limits_{k=2}^{\infty}
\frac{\alpha^{k-1}{\Omega_{1}^{{\rm{k}}-{1}}\Omega_{2}^{1}}}{k!}+{\Omega}_{2}
\end{align}
where $\underset{(M)}{=}$ denotes the neglect of terms of the form 
i) $\Omega_{1}^{M+1}$, 
ii) $\Omega_{1}^{M}\Omega_{2}^{1},{\Omega }_{1}^{M-1}{\Omega }_{2}^{1}{\Omega }_{1},\ldots$, 
iii) ${\Omega}_{2}^{k}$ for $k\geq 2$, and 
iv) $\Omega_p$ for $p\geq 3$. 
By construction, after applying Taylor KIK mitigation, all terms proportional to $\alpha_k$ for $1\leq k\leq M$ vanish and we get
\begin{align}
\label{eq:27}
K_{mit}^{(M)}\underset{(M)}{=}U(I+{\Omega }_{2}).
\end{align}
This $\Omega_2$ term constitutes the leading order in the bias of the global KIK method. 
The limit $M\rightarrow\infty$ leads to the GKIK formula (Eq.~\eqref{eq: GKIK formula}). Next, without loss of generality, we study three layers by Taylor expanding each layer. The analysis proceeds similarly as in the single-layer case. 
The mitigated operator now reads
\begin{align}
{K}_{mit}^{(M)}\mathop{=}\limits_{(M)}\mathop{\sum }\limits_{j=0}^{M}{a}_{j,Tay}^{(M)}{U}_{c}{U}_{b}{U}_{a}\left[\mathop{\sum }\limits_{k=0}^{M}\left[\mathop{\sum }\limits_{{k}_{c}+{k}_{b}+{k}_{a}=k}{(2j+1)}^{k}\frac{{\Omega }_{1,c}^{{k}_{c}}{\Omega }_{1,b}^{{k}_{b}}{\Omega }_{1,a}^{{k}_{a}}}{{k}_{c}!{k}_{b}!{k}_{a}!}\right]+\mathop{\sum }\limits_{l=a,b,c}{\Omega }_{2,l}\right]={U}_{tot}\left(I+\mathop{\sum }\limits_{l=a,b,c}{\Omega }_{2,l}\right),
\end{align}
where ${\Omega }_{2}^{LKIK}={\sum }_{l=a,b,c}{\Omega }_{2,l}$ is the analog of $\Omega_2$ in the GKIK expression Eq.~\eqref{eq:27}. 
By comparing ${\Omega }_{2}^{LKIK}$ and $\Omega_2$, we observe that the local contribution to $\Omega_2$ from each layer is the same (blue triangles in Fig.~\ref{fig: illust omega2}), but ${\Omega }_{2}^{LKIK}$ has no cross-layer $\Omega_1$ contributions. 
As explained later, since only the triangular regions contribute to ${\Omega}_{2}^{LKIK}$, and their contribution becomes negligible as the number of layers increases, the error bias becomes negligible as well. This is the basis for the second key result in this paper: LKIK is bias-free when the number of layers is sufficiently large.

From a practical point of view, by bias-free we mean that for any target accuracy, one can choose $L$ such that the 
$\sum_{l=1}^{L}\tilde{\Omega}_{2,l}$
term produces a bias which is much smaller than the target accuracy. 
We emphasize that $\Omega_{2}^{LKIK}$ is the effective $\Omega_2$ term after carrying out mitigation with Taylor coefficients. 
The circuit associated with a certain amplification level still has contributions from the square regions. 
For $M\to\infty$ and to leading order in $\Omega_2$, we can write
\begin{align}
\label{eq: Kinf omega2 prod}
K_{mit}^{(\infty)} & \cong  {U}_{tot}(I+\mathop{\sum}
\limits_{l=a,b,c}{\widetilde{\Omega }}_{2,l})
\\ & \cong  {U}_{c}(I+{\Omega }_{2,c}){U}_{b}(I+{\Omega }_{2,b}){U}_{a}(I+{\Omega }_{2,a}).\nonumber
\end{align}
Using Eqs.~(\ref{eq: GKIK formula}) and (\ref{eq: Kinf omega2 prod}) we obtain the Layered-KIK formula
\begin{equation}
\label{eq: Layered KIK formula}
K_{mit\:LKIK}^{(\infty)}=K_{L}\frac{1}{\sqrt{K_{L}^{I}K_{L}}}...K_{i}\frac{1}{\sqrt{K_{i}^{I}K_{i}}}...K_{1}\frac{1}{\sqrt{K_{1}^{I}K_{1}}}.
\end{equation}
In Fig.\ref{fig:A-four-qubit-simulation}, the global KIK asymptote in Eq.~\eqref{eq: GKIK formula} is shown in the red dashed line. 
The other dashed lines show the LKIK asymptotes \eqref{eq: Layered KIK formula} for each number of layers. 
This demonstrates that Eq.~\eqref{eq: Layered KIK formula} correctly describes the behaviour of the Layered KIK scheme at high mitigation order. 
Figure \ref{fig:A-four-qubit-simulation} also illustrates how the Layered KIK mitigation becomes bias-free as $L$ increases.

\subsubsection{The residual error scaling as a function of the number of layers}

In this section, we quantitatively study the scaling behaviour of the vanishing bias term as a function of the number of layers. 
First, we obtain an upper bound that is applicable to any partition, and then study the exact scaling in the limit of thin layers.
We start by expressing the bias size using the operator norm
\begin{align}
\nrm{U-K_{mit}^{(\infty)}}_{op} & \cong\nrm{\sum_{l=1}^L\tilde{\Omega}_{2,l}}_{op}\le\sum_{l=1}^L\nrm{\Omega_{2,l}}_{op}\nonumber\\
&\le\sum_{l=1}^L\frac{1}{2}\intop_{t_{l}}^{t_{l+1}}dt\intop_{t_{l}}^{t}dt'\nrm{[\mc L_{int}(t),\mc L_{int}(t')]}\nonumber\\
&\le\frac{1}{2}\sum_{l=1}^L(t_{l+1}-t_l)^2\nrm{\mc L(t_{max})}_{op}^{2},
\end{align}
where $t_{max}$ is the time that maximizes $\nrm{\mc L(t)}_{op}$.
To make further progress, we assume that the layers are uniform in
width such that $t_{l+1}-t_{l}=\tau/L$ and we get
\begin{equation}
\label{eq: bias upper bound}
\nrm{U-K_{mit}^{(\infty)}}\le\frac{\tau^{2}}{2L}\nrm{\mc L(t_{max})}_{op}^{2}
\end{equation}
which clearly vanishes for a large $L$, since both $\tau$ and $\parallel{\mathcal{L}}({t}_{max})\parallel$ are independent of $L$. 
To obtain Eq.~\eqref{eq: bias upper bound}, we have used 
$\parallel [{{\mathcal{L}}}_{int}(t),{{\mathcal{L}}}_{int}(t{\prime} )]\parallel \le \parallel {{\mathcal{L}}}_{int}(t){{\mathcal{L}}}_{int}(t{\prime} )\parallel +\parallel {{\mathcal{L}}}_{int}(t{\prime} ){{\mathcal{L}}}_{int}(t)\parallel$
 and the sub-multiplicativity of the operator norm ${\parallel {\mathcal{L}}(t){\mathcal{L}}(t{\prime} )\parallel }_{op}\le {\parallel {\mathcal{L}}(t)\parallel }_{op}{\parallel {\mathcal{L}}(t{\prime} )\parallel }_{op}$. 
For norms that do not satisfy this, ${\parallel {L}_{int}({t}_{max})\parallel }^{2}$ should be replaced by ${\parallel {\mathcal{L}}({t}_{max1}){\mathcal{L}}({t}_{max2})\parallel }_{op}$. 
The approximate uniformity of the layers is important for obtaining the suppression of the bias. 
Imagine one layer which is almost equal to the whole interval $\tau$, with many very thin layers filling the remaining gap. 
In such a case, the bias will be dominated by the thick layer, and no residual bias suppression will be achieved compared to GKIK. 
The derivation above involves multiple applications of inequalities so it is possible that the actual bias suppression is even stronger than Eq.~\eqref{eq: bias upper bound}. 
In Supplementary Note \ref{SupplementaryNote6}, we show that significant deviations from uniformity across the layers only slightly affect the bound in Eq.~\eqref{eq: bias upper bound}, and it retains the $1/L$ scaling.

\begin{figure}
\centering
\includegraphics[width=8cm]{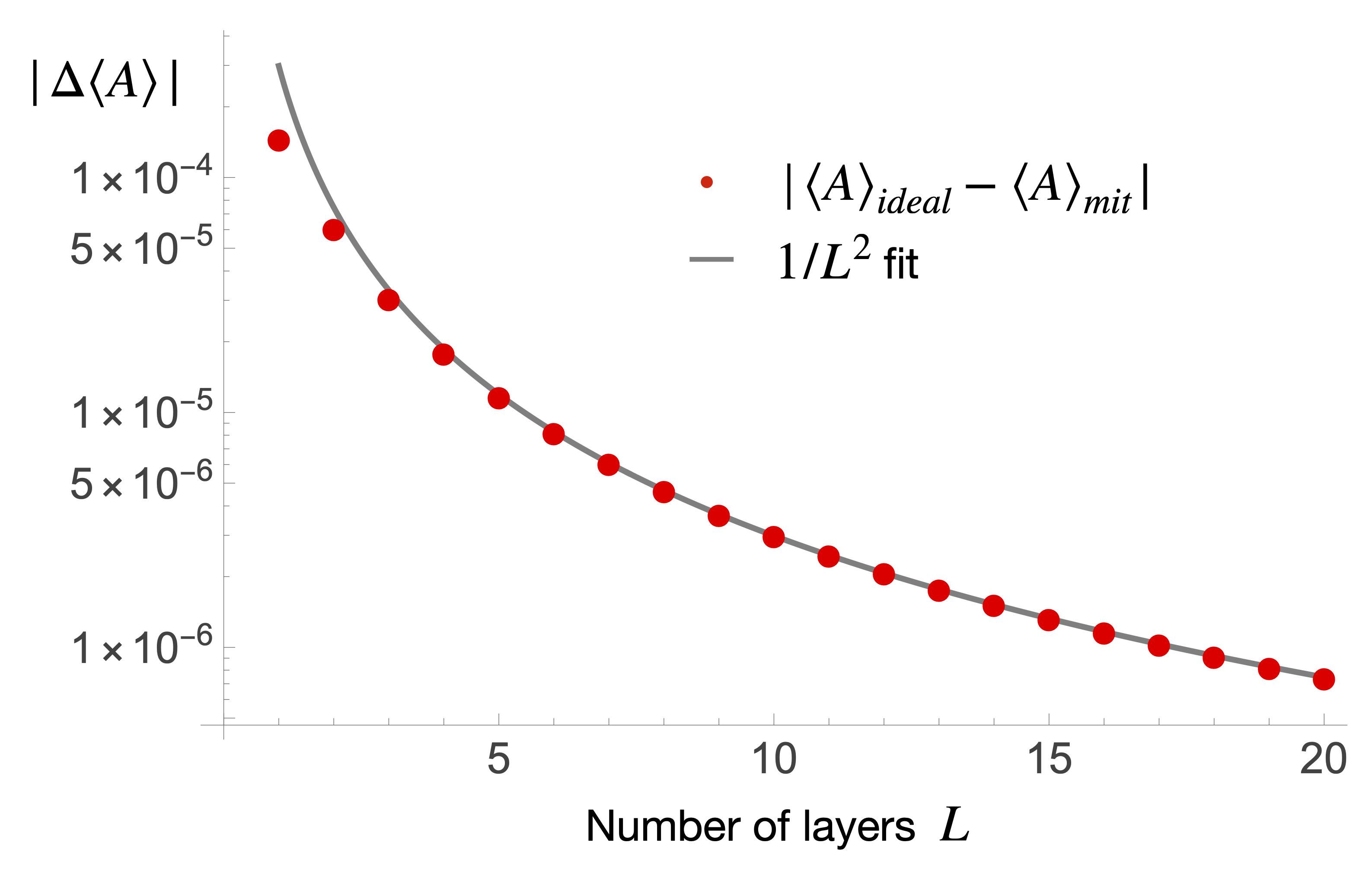}
\caption{\label{fig: FF}For the same example as in Fig.~\ref{fig:A-four-qubit-simulation}a, the error with respect to the ideal expectation value (red dots) is plotted for order seven as a function of number of layers. 
The good fit to the $1/L^2$ curve confirms the prediction of Eq.~\eqref{eq:34}.}
\end{figure}

Interestingly, a tighter yet still rather generic error bound can be obtained in the limit of thin, uniformly-spaced layers such that $t_{l+1}-t_{l}=\frac{\tau}{L}\ll\tau$. 
To proceed, we make the following assumptions: 
i) The layers are sufficiently thin, such that in each layer the Hamiltonian and the dissipator can be considered as time-independent. 
ii) Since the layers are thin, $\nrm{\mc H(t)}_{op}\tau/L\ll1$. 
By expanding in $\tau/L$, we obtain
\begin{equation}
\Omega_{2,l}=\frac{1}{3}(\tau/L)^{3}[\mc L,[H,\mc L]].
\end{equation}
After summing over all the layers, the effective accumulated $\Omega_{2}$
correction is
\begin{align}
\Omega_{2}^{eff} &= 
\frac{\tau^2}{3L^2}\sum_{l}[\mc L(t_{l}),[H(t_{l}),\mc L(t_{l})]](\tau/L)\nonumber\\
\label{eq:34}
&\to\frac{\tau^2}{3L^2}
\intop_{0}^{\tau}dt[\mc L(t),[H(t),\mc L(t)]],
\end{align}
where we have used the fact that for a sufficiently large number of layers, the sum can be approximated by an integral. 
Since the integral does not depend on $L$, it follows that the bias correction decays as $1/L^2$, which is substantially faster than the coarse bound given in Eq.~\eqref{eq: bias upper bound}. 
To demonstrate the $1/L^2$ dependence in thin layers, in Fig.~\ref{fig: FF} we revisit the example of Fig.~\ref{fig:A-four-qubit-simulation}a and this time plot the error for order seven as a function of the number of layers. 
Order seven was chosen because it ensures that errors associated with powers of $\Omega_1$ is sufficiently mitigated, and the error is dominated by the $\Omega_2$ term. 
The gray line represents a fit to $1/L^2$ using the last data point, i.e. $\Delta \langle A\rangle_{L=20}(20)^2/L^2$.
The good fit confirms the predicted $1/L^2$ error dependence for thin layers.

While finalizing the writing of this paper, we learned that layer-based amplification has been applied using multi-variable extrapolation ZNE \cite{Mari2024LayerBasedRich}. 
Despite a clear overlap in the conceptual idea of amplifying the noise layer by layer, the results obtained in ref.~\cite{Mari2024LayerBasedRich} and in the present work differ significantly. 
In our approach, the amplification protocol of each layer, ${K}_{l}{({K}_{l}^{I}{K}_{l})}^{m}$, is based on the pulse-inverse \cite{npjqiKIK} while in ref.~\cite{Mari2024LayerBasedRich} it is mathematically represented using the Hermitian conjugate of the original channel, i.e. ${K}_{l}{({K}_{l}^{\dagger }{K}_{l})}^{m}$. 
This distinction is crucial, since an accurate and simple way for implementing $K^\dagger$ is, to the best of our knowledge, not known for cases where the noise mechanism is nontrivial and the noise parameters are unknown (noise characterization is not allowed, since we aim for drift-resilient methods).

When using the operational protocol $K^I$ instead of $K^\dagger$, one must account for high-order Magnus terms, as is done in the present paper. 
Alternatively, choosing the gate-insertion method \cite{he2020zero} leads to incorrect amplifications in the common case where the noise does not commute with the unitary. Therefore, an explicit analysis of the amplification method is crucial. 
In our case, this analysis shows that LKIK systematically suppresses high-order corrections resulting from the use of the operationally accessible $K^I$ instead of the mathematical $K^\dagger$. 
We further discuss the runtime cost of \cite{Mari2024LayerBasedRich} in section ``The Layered KIK formula and its relation to other noise amplification schemes''.

Moreover, our approach guarantees convergence to the correct result, as each order of mitigation eliminates a high-order noise term. 
This is also supported by the asymptotic expression for infinite mitigation order \eqref{eq: Layered KIK formula}. 
Finally, we show that LKIK is consistent with mid-circuit measurement by leveraging properties of the Taylor coefficients. 
It is unclear whether the coefficients obtained from multi-variate Richardson extrapolation can be used for showing compatibility with MCM. 
That being said, the work \cite{Mari2024LayerBasedRich} is very interesting and it is intriguing to check whether 
1) it becomes bias-free in the limit of a large number of layers and 
2) it is compatible with dynamic circuits.

\begin{figure}
\centering
\includegraphics[width=8cm]{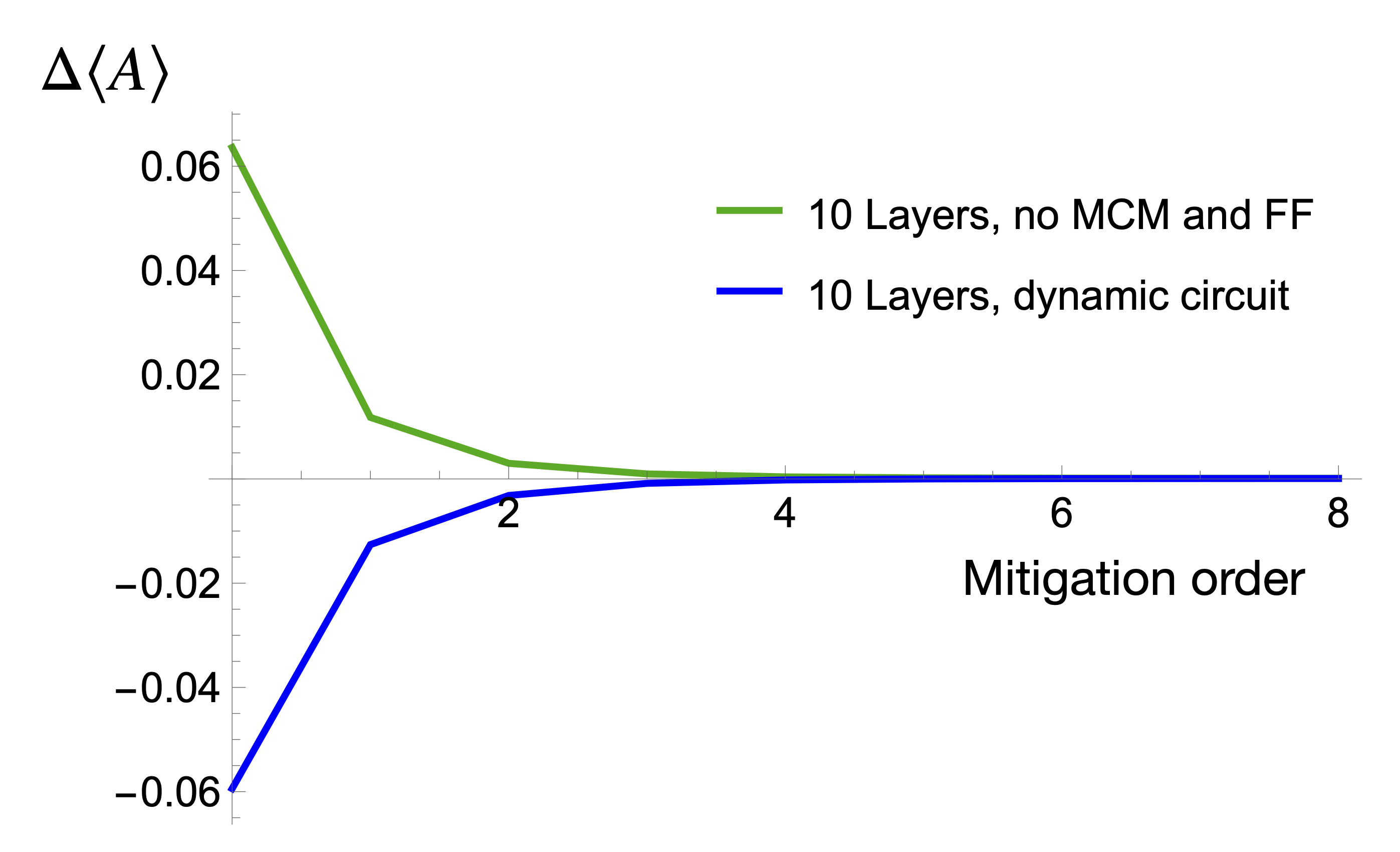}
\caption{\label{fig:fig4}Demonstration of LKIK mitigation of a dynamic circuit (mid-circuit measurement and feedforward, blue curve) and a unitary evolution (green curve).The circuit is the same as in Fig.~\ref{fig:A-four-qubit-simulation} (ten-layer case). 
The feedforward is applied after each layer; see text for details. As expected from theory, LKIK performs equally well in both scenarios.}
\end{figure}

A numerical simulation that illustrates LKIK mitigation in a dynamic circuit is presented in Fig.~\ref{fig:fig4}. 
The green curve serves as a reference showing LKIK mitigation of a non-dynamic circuit as in Fig.~\ref{fig:A-four-qubit-simulation} with a noise parameter of $\xi = 0.1$ and $L = 10$ layers. 
The blue curve is the outcome of a simulation with the same parameters but with MCM and feedforward, i.e. a dynamic circuit. 
After each layer, the first qubit is measured in the computational basis. 
If the outcome is ‘0’, no action is taken; if the outcome is ‘1’, Hadamard gates are applied to qubits 2–4. 
While the feedforward substantially changes the result, $\Delta \langle A\rangle \to 0$ in both cases.

\subsubsection{Compatibility with post-selection}

On top of mid-circuit measurements and feedforward, dynamic circuits may also contain post-selection where a subset of the outcomes is kept and the rest are discarded. 
Post-selection is used, for example, for implementing weak measurements.

For concreteness, consider a mid-circuit measurement on qubit number one and suppose that the quantity of interest is the expectation value of an observable $A$ given that the mid-circuit outcome s is ‘0’. 
Denoting by ${\mathbb{M}}$ the projector to this state, we have
\begin{equation}
\label{eq:35}
\langle A|s='0'\rangle = \frac{\langle O|K_b \mathbb{M} K_a|\rho_0\rangle}{P(s='0')} = \frac{\langle O|K_b \mathbb{M} K_a|\rho_0\rangle}{\langle \mathbb{M}|K_a|\rho_0\rangle},
\end{equation}
where $K_a(K_b)$ is the noisy evolution operator of the layer before (after) the mid-circuit measurement. 
The ideal expression is 
\begin{align}
{\langle A| s{=}^{{\prime} }{0}^{{\prime} }\rangle }_{ideal}=\frac{\langle O| {U}_{b}{\mathbb{M}}{U}_{a}| {\rho }_{0}\rangle }{\langle {\mathbb{M}}| {U}_{a}| {\rho }_{0}\rangle }.
\end{align}
We suggest applying LKIK to mitigate \eqref{eq:35} using two mitigation processes: one for the numerator and one for the denominator. 
$\langle O| {K}_{b}{\mathbb{M}}{K}_{a}| {\rho }_{0}\rangle$ and $\langle {\mathbb{M}}|{U}_{a}|{\rho }_{0}\rangle$ are both dynamic circuits without post-selection that can be mitigated using the LKIK approach as described earlier.

\subsection{The Layered KIK formula and its relation to other noise amplification schemes}

The Layered KIK formula can represent the asymptotic performance of different mitigation schemes. For example, the Layered KIK formula can also arise from separate mitigation of each layer
\begin{align}
\label{eq:37}
K_{mitigate\ layers}^{(m)} = K_L^{(m)} \dots K_2^{(m)} K_1^{(m)},
\end{align}
here ${K}_{l}^{(m)}$ is the $m-$th order Taylor mitigation of layer $l$. 
When $m\rightarrow\infty$ the Layered KIK formula is obtained. 
For a two-layer circuit with first-order Taylor mitigation, the product of the mitigated layers reads
\begin{align}
\label{eq:38}
{K}_{b}^{(1)}{K}_{a}^{(1)} & =  (\frac{3}{2}{K}_{b}-\frac{1}{2}{K}_{b}{K}_{b}^{I}{K}_{b})(\frac{3}{2}{K}_{a}-\frac{1}{2}{K}_{a}{K}_{a}^{I}{K}_{a})\nonumber\\ 
& =  \frac{9}{4}{K}_{b}{K}_{a}-\frac{3}{4}{K}_{b}{K}_{b}^{I}{K}_{b}{K}_{a}-\frac{3}{4}{K}_{b}{K}_{a}{K}_{a}^{I}{K}_{a}+\frac{1}{4}{K}_{b}{K}_{b}^{I}{K}_{b}{K}_{a}{K}_{a}^{I}{K}_{a}.
\end{align}
While this scheme provides powerful mitigation, its sampling overhead grows exponentially with the number of layers. 
For example, for first-order mitigation of ten layers the sampling overhead will be approximately one million ($2^{20}$). 
Note that compared to the scheme described earlier, there are circuits, i.e. terms in Eq.~\eqref{eq:38}, where different layers are amplified by different factors.

Next, we suggest an alternative scheme with a reduced sampling overhead, that can be derived from Eq.~\eqref{eq: Layered KIK formula} using the definition 

\begin{align}
\epsilon_{i}={K}_{i}^{I}{K}_{i}-I,
\end{align}
and then Taylor expanding the expression
\begin{align}
\label{eq:40}
{K}_{L}\frac{1}{\sqrt{I-{\epsilon }_{L}}}\ldots {K}_{l}\frac{1}{\sqrt{I-{\epsilon }_{l}}}\ldots {K}_{1}\frac{1}{\sqrt{I-{\epsilon }_{1}}}
\end{align}
as a multi-variable function. 
For example, for first-order mitigation, only the terms linear in each $\epsilon$ will be kept.
As a result, the noise amplification circuits will have the form of ${\sum }_{l=1}^{L}{K}_{L}\ldots ({K}_{l}{K}_{l}^{I}{K}_{l})\ldots {K}_{1}$, which means that only one layer is amplified in each circuit (a term in this sum). 
Such amplification circuits are shorter than the circuits used in Layered KIK with Taylor coefficients of the same order, but the sampling overhead is higher.

The first-order expansion leads to the same result obtained from the NOX framework when the layer noise amplification is set to three. 
While the NOX framework also includes non-integer amplifications of each layer, it is still a first-order mitigation method that does not cancel second-order terms in the noise, as can be verified by explicit Taylor expansion of the NOX-mitigated operator.

The procedure based on multi-variable Taylor expansion of \eqref{eq:40} provides a systematic way to construct high-order mitigation protocols.
A second-order mitigation, based on multi-variable second-order expansion of \eqref{eq:40}, will suppress the ${\epsilon }_{l}^{2}$ terms and the ${\epsilon }_{{l}_{2}}{\epsilon }_{{l}_{1}}$ terms where $l_2 > l_1$. 
After some algebra, we find that second-order mitigation in layers takes the form
\begin{align}
{K}_{mitigate\,layers}^{(2)}&=(1+\frac{L(L+6)}{8}){K}_{L}{K}_{L-1}\ldots {K}_{1}-(L+\frac{{L}^{2}}{4}){S}_{1}+\frac{3L}{8}{S}_{2}+\frac{L(L-1)}{8}{S}_{3},
\\
{S}_{1}&=\frac{1}{L}[{K}_{L}^{[3]}{K}_{L-1}{K}_{L-2}\ldots {K}_{1}+{K}_{L}{K}_{L-1}^{[3]}{K}_{L-2}\ldots {K}_{1}],
\\
{S}_{2}&=\frac{1}{L}[{K}_{L}^{[5]}{K}_{L-1}{K}_{L-2}\ldots {K}_{1}+{K}_{L}{K}_{L-1}^{[5]}{K}_{L-2}\ldots {K}_{1}],
\\
{S}_{3}&=\frac{1}{L(L-1)/2}[{K}_{L}^{[3]}{K}_{L-1}^{[3]}{K}_{L-2}\ldots {K}_{1}+{K}_{L}^{[3]}{K}_{L-1}{K}_{L-2}^{[3]}\ldots {K}_{1}+\cdots +{K}_{L}{K}_{L-1}\ldots {K}_{2}^{[3]}{K}_{1}^{[3]}],
\end{align}
where we used the notation ${K}_{l}^{[m]}={K}_{l}{({K}_{l}^{I}{K}_{l})}^{\frac{m-1}{2}}$ for the $m-$th amplification factor of a layer $l$. 
The sampling overhead of this method is $1+\,\frac{L(L+4)}{2}$, while for the linear theory it is $1+\frac{L+1}{2}$. 
We observe that by restricting to lower-order error ${\Omega }_{1}^{3}$ in this case), the overhead can be polynomial in the number of layers, rather than exponential as in the case of a mitigated operators product \eqref{eq:37}. 
However, this reduction comes at the cost of reduced performance, compared to full multiplication of the mitigated evolution operators for each layer. Interestingly, the multivariate extrapolation (MVE) QEM scheme introduced and analyzed in ref.~\cite{Mari2024LayerBasedRich} leads to precisely the same coefficients as those obtained from the multi-variable Taylor expansion outlined above. Since the MVE coefficients were already derived in ref.~\cite{Mari2024LayerBasedRich}, the main additional contribution of the present work (in this setting) is the insight that increasing the number of layers allows one to suppress the $\Omega_2$ bias that arises when using experimentally realistic amplification protocols. In addition, the Layered KIK formula \eqref{eq: Layered KIK formula} characterizes the remaining bias for any fixed, finite number of layers in the limit of infinite mitigation order.

It is important to note that the MVE differs from the single-layer Taylor (SLT) coefficient with LKIK as used in the previous sections. 
In particular, in SLT, all layers are amplified by the same factor. 
Consequently, the average depth of the circuits in mitigation order $m$ is approximately $m$ times longer than the original circuit. 
The overall runtime overhead is the product of the sampling overhead and the depth overhead. 
In Supplementary Note \ref{SupplementaryNote7}, we investigate the runtime overhead and discover that, although the MVE depth overhead is smaller than that of the SLT, this advantage is negligible with respect to the significantly larger sampling overhead of the MVE. 
Consequently, we conclude that the runtime overhead is greater for the MVE. 
For detailed information regarding the performance of both methods and methods for comparing them, please refer to Supplementary Note \ref{SupplementaryNote7}.

That being said, when using SLT without employing Layered KIK, i.e., using GKIK, the comparison becomes more nuanced. 
Using MVE with more layers provides stronger mitigation compared to GKIK, as MVE effectively addresses the $\Omega_2$ bias.

\section{Discussion}

In this work, we have presented the first, to the best of our knowledge, error mitigation framework for dynamic circuits that is both drift-resilient and bias-free. 
As such, Layered KIK (LKIK) can be applied to any dynamic circuit, including quantum error correction codes. 
Since quantum error correction is not expected to work perfectly for any type of error and for any depth, this work paves the way to reliable and drift-resilient error mitigation of circuits that undergo imperfect quantum error correction. Therefore, on platforms compatible with KIK amplification, any achievement of quantum error correction codes can be boosted by applying LKIK. An experimental realization of LKIK in a dynamic circuit has been presented in a companion paper \cite{2024E2Eparity}.

As explained in the introduction, this seamless integration with QEC is a major advantage over popular error mitigation methods that rely on error characterization, such as probabilistic error cancellation (PEC), probabilistic error amplification (PEA), tensor error mitigation (TEM), and more. LKIK is also compatible with post-selection, which enables the use of syndrome measurements to post-select and screen out specific uncommon errors that would otherwise substantially increase the error mitigation overhead. The above claims about QEC-LKIK integration assume that the mid-circuit measurement errors are negligible. If this assumption does not hold, it is possible to use PEC for measurements as suggested and studied in refs.~\cite{hashim2025PECreadout,Temme2024dynamicPEC,koh2026readout}. 
A drift-resilient solution that does not require readout errors characterization has been presented and demonstrated experimentally in ref.~\cite{2024E2Eparity}.

In trapped ions quantum computers, this work is already applicable as small-scale error correction experiments have begun to emerge \cite{quantinuum2024_12LogicalQubits,quantinuum4dsurface,Quantinuum20243qQFTsteane,Quantinuum2024adder1bitFT,Quantinuum2024betterthanphysical,Quantinuum2024FTteleport,huang2024comparing,Quantinuum2022_5qcolor,egan2021SingleIonQEC,Quantinuum2021realizationQEC}. 
Since this platform is compatible with LKIK, it should be possible to observe the advantage of combining LKIK with QEC using currently available devices. This advantage is expected to persist for larger circuits with error correction once they can be implemented. The KIK method presumes time-dependent Markovian noise; however, in actual quantum devices, non-Markovian effects can be significantly pronounced. A common source of non-Markovianity is crosstalk to spectator qubits. Fortunately, this and other types of non-Markovian noise can be efficiently addressed using dynamical decoupling, Pauli twirling (see Supplementary Fig.~9b in ref. \cite{npjqiKIK}), or pseudo twirling (see Fig.~6 in ref. \cite{santos2024pseudo}) before applying the KIK protocol. For methods of characterizing and addressing non-Markovian effects, refer to refs.\cite{PhysRevA.105.022428,PhysRevX.15.021047,PRXQuantum.6.030202,White2020,PhysRevApplied.17.054018,Ahsan2023-sk,10.5555/3370214.3370216}.

The Layered KIK protocol does not introduce additional hardware or time overhead compared to the original global KIK protocol. However, the more frequent sign changes of the drive may necessitate faster control electronics. We anticipate that if the layers are not narrower than a single qubit gate, the existing control electronics should be sufficient. For thinner layers, the potential for leakage noise and excitations of non-comptational states should be mitigated through pulse shaping techniques. Finally, we point out that LKIK is not an exclusive framework. It can be combined with other mitigation frameworks to gain additional advantage. For example, PEC, PEA and TEM may be more efficient in terms of shot count compared to LKIK, but they lack drift resilience. By applying TEM, for example, and then LKIK, the outcome will be noise-drift resilient and the sampling overhead will be smaller than that of using LKIK alone.

\section{Methods}

\subsection{The KIK drift resilience and the importance of pulse inverse}

The first experiment reported in this manuscript is described in Supplementary Note \ref{SupplementaryNote1}. 
The experiment was performed on the trapped ion quantum computer IBEX (Alpine Quantum Technology - AQT). 
The goal of this experiment was to demonstrate the drift resilience of the KIK method. 
To this end, we artificially induced two different over-rotations in a circuit composed of four repetitions of the $R_{xx}(\pi/2)$ gate. 
In the first half of the experiment, the rotation was $\pi/2-0.3$, while in the second half it was $\pi/2-0.5$, where the ideal angle is $\pi/2$. 
This error profile mimics an abrupt and unpredictable drift in the noise. 
After applying Pauli twirling, this coherent error was converted into an incoherent error. 
Although only two qubits were used for the computation, we applied Pauli twirling to all eight qubits loaded in the trap, to suppress crosstalk that gives rise to non-Markovian effects. 
More information about the experiment can be found in Supplementary Note \ref{SupplementaryNote1}.

The second experiment is described in Supplementary Note \ref{SupplementaryNote2}. 
This experiment was performed on the device \textit{ibm\_jakarta} from IBM. 
The goal of the experiment was to demonstrate that the incorrect noise amplification generated by gate insertion can lead to substantial errors and nonphysical results. 
In this two-qubit experiment, our circuit consists of a sequence of ten swap gates, and we measure the survival probability with respect to the initial state $|11\rangle$.
More details about the experiment, including the results, can be found in Supplementary Note \ref{SupplementaryNote2}.

\begin{acknowledgments}
Raam Uzdin is grateful for support from the Israel Science Foundation
(Grants No. 2556/20 and 2724/24). The support of the Israel Innovation
Authority is greatly appreciated.
\end{acknowledgments}


 \bibliographystyle{apsrev4-2}
 \bibliography{Refs_LKIK}


\pagebreak
\newpage

\appendix


\setcounter{section}{0}
\renewcommand{\thesection}{\arabic{section}}
\titleformat{\section}{\normalfont\bfseries}{Supplementary Note \thesection \ -}{1em}{}

\setcounter{figure}{0}
\renewcommand{\figurename}{Supplementary Figure}
\renewcommand{\thefigure}{\arabic{figure}}

\setcounter{table}{0}
\renewcommand{\tablename}{Supplementary Table}
\renewcommand{\thetable}{\arabic{table}}

\counterwithout{equation}{section}   
\setcounter{equation}{0}
\renewcommand{\theequation}{S\arabic{equation}}

\onecolumngrid

\section{\hspace{-2mm}Trapped Ion Experiment Demonstrating the KIK Drift Resilience}
\label{SupplementaryNote1}

To demonstrate the drift resilience of the KIK method, we have conducted an experiment using the Alpine Quantum Technology (AQT) trapped ion quantum computer IBEX. 
In this experiment, we evaluated the robustness of the KIK method against errors whose parameters drifted substantially
during the experiment. 
To control the noise parameters and their drift, we induced noise in the following way. 
We conducted two consecutive experiments with two different over-rotation coherent errors. 
Upon applying randomized compiling (RC) to the gates the coherent error was converted into an incoherent error whose intensity was determined by the over-rotation. 

We used a circuit consisting of four repetitions of the $R_{xx}(\pi/2)$ gate. 
We used only two qubits for the computation but applied Pauli twirling to all eight qubits loaded to the trap, to remove crosstalk that gives rise to non-Markovian effects.
This circuit is logically equivalent to an identity gate.
The expectation value we measured is the survival probability, which corresponds to the final population of the initial state $\ket{00}$. 
Single-qubit Pauli gates were sampled for implementing the Pauli twirling (randomized compiling).
The coherent error was introduced through a mis-rotation in the $R_{xx}$ gate: in the first half of the experiment the rotation was $\pi/2-0.3$ and in the second half it was $\pi/2-0.5$ where the ideal angle is $\pi/2$. 
This error profile mimics an abrupt and unpredictable drift
in the noise.

We compared two different circuit execution orders.
The first execution order which guaranties drift-resilience involves the following operations: execution of a small number of shots - twenty in the present case - without noise amplification, then an execution of the same number of shots with an amplification factor of three, then another execution with amplification factor five, and so on.
The sequence is repeated with different random choice of Pauli twirling gates in each round, until the desired final accuracy is achieved. 
Two hundred rounds were used in this experiment. 

The other execution order executes first $20\times200=4$k
shots for the unamplified circuit, then $4$k shots for an amplification factor of three, and finally $4$k shots for an amplification factor of five. 
It is convenient to quantify the difference between the two
execution orders using a hopping parameter. 
In the first case, the hopping parameter is twenty shots, i.e., there is a hopping to a circuit with a different amplification level every twenty shots. 
In the other case (no hopping) the hopping parameter is 4k, which means there is no hopping between circuits; 
the shots of the next amplification level will start only after finishing the previous amplification level.
See also Fig.~2 in the Supplementary Information of Ref.~\cite{npjqiKIK} for an illustration of the two execution orders.

\begin{figure}[ht!]
\centering
\includegraphics[width=0.7\linewidth]{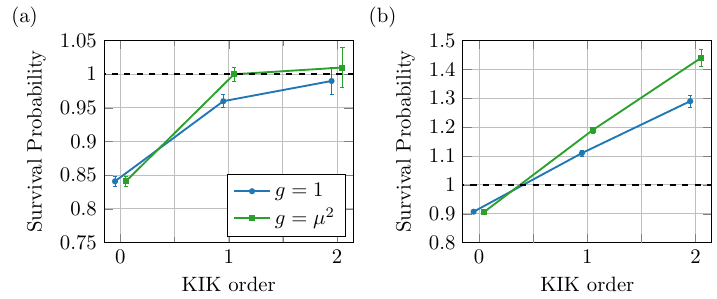}
\caption{\label{fig: AQT}Experimental results from the AQT trapped ion quantum computer IBEX. 
The plots show mitigation of time-dependent
error using (a) drift-resilient execution order and (b) non drift-resilient execution order. 
The time-dependent error is injected by using randomize
compiling to convert a controllable time-dependent coherent error
into a controllable time-dependent incoherent error. 
See text for the description of the coherent error time-dependence. 
In (a), a different level of noise amplification is executed every twenty shots, so that a full cycle of the KIK algorithm is executed before the noise changes. 
In (b), 4k shots are taken sequentially for each of the amplification factors. The results show that protocol (b) leads to unphysical results (survival probability beyond one) while protocol (a) converges to the correct result (dashed line). 
The blue curves exploit the Taylor coefficients while the green curve shows mitigation using the adaptive coefficient described in Ref.~\cite{npjqiKIK}.}
\end{figure}

\begin{table}[ht!]
\caption{\label{tab:aqt_data}Experimental results from the AQT trapped ion quantum computer IBEX. The data presented in Tab.(a) and (b) are the same used to in Supplementary Figure~\ref{fig: AQT}.}
\centering
\begin{minipage}[t]{0.45\textwidth}
\centering
{\small Tab. (a): Drift-resilient execution order.}
\begin{tabular}{cccc}
\toprule
order & Unmitigated& $g=1$ & $g=\mu^2$ \\
\hline
0 & 0.841(7) & 0.841(7)  & 0.841(7) \\
1 & 0.61(1) & 0.96(1) & 1.00(1) \\
2 & 0.48(1) & 0.99(2) & 1.01(3) \\
\hline
\end{tabular}
\end{minipage}
\hfill
\begin{minipage}[t]{0.45\textwidth}
\centering
{\small Tab. (b): Non drift-resilient execution order.}
\begin{tabular}{cccc}
\toprule
order & Unmitigated& $g=1$ & $g=\mu^2$ \\
\hline
0 & 0.907(5) & 0.907(5)  & 0.907(5) \\
1 & 0.50(2) & 1.11(1)  & 1.19(1) \\
2 & 0.57(1) & 1.29(2)  & 1.44(3) \\
\hline
\end{tabular}
\end{minipage}
\end{table}

Supplementary Figure~\ref{fig: AQT} compares the drift-resilient, twenty-shot hopping execution order (a), to drift-sensitive, no-hopping execution order (b) (the corresponding data is presented in Supplementary Table~\ref{tab:aqt_data}. 
The dashed line shows the ideal result (survival probability
equals one for an ideal identity circuit). 
The twenty-shot hopping execution order converges to the correct result despite the abrupt change in the noise parameters in the middle of the experiment. 
In contrast, the no-hopping execution order leads to unphysical results which greatly exceed one. 
This is due to the fact that noise parameter are different for each amplification circuit.

One can think of the drift resilience in the following way: a complete execution of the KIK protocol (in particular the LKIK) removes the bias when the noise parameters remain fixed in time. The noise parameters can still change from one gate type to another, but each time a specific gate is used, its noise parameters are the same. 
In principle, this bias removal holds even if a single shot is used for each circuit.
Let $x_{0}$ be the ideal expectation value and $d(t)$ be some time-dependent drift. 
The variance of the noiseless observable is $\sigma_{0}$ and
the sampling overhead is $\gamma$. 
If the hopping parameter is sufficiently small, we can assume that during the $i$-th execution round $d(t)$ stays fixed, and therefore the unmitigated values in each round $x_{i}$ will have an expectation value $\mu(x_{i})=x_{0}+d(t_{i})$ and a standard deviation of $\sigma(x_{i})=\sigma_{0}/\sqrt{N_{\text{hop}}}$.
However, after mitigation we get
\begin{align}
\mu(x_{i}^{mit}) & =x_{0},\\
\sigma(x_{i}^{mit}) & =\sigma_{0}\gamma/\sqrt{N_{\text{hop}}}.
\end{align}
Crucially, the mitigation removes the bias from the expectation value.
The statistics of the mean over $N_{\text{rounds}}$ rounds is
\begin{align}
\mu\left(\frac{1}{N_{rounds}}\sum_{i}x_{i}\right) & =x_{0},\\
\sigma\left(\frac{1}{N_{rounds}}\sum_{i}x_{i}\right) & =\sigma_{0}\gamma/\sqrt{N_{\text{hop}}N_{\text{rounds}}} ,
\end{align}
where the last result follows from the statistical independence of the different rounds.

\section{\hspace{-2mm}Pulse Inverse (KIK) vs gate insertion experiment}
\label{SupplementaryNote2}

In Ref.~\cite{npjqiKIK} it was shown analytically that gate insertion (GI) is incorrect in all orders yielding a mitigated evolution operator of the form
\begin{equation}
K_{mit}^{GI}=U(I+\frac{1}{2}[U,N]),
\end{equation}
where $N$ is the noise channel and $U$ is the ideal CNOT or CPHASE evolution operator. 
Thus, GI is only valid if $U$ commutes with the
noise $N$. 
For instance, in cross-resonance gates CNOT gates, $U$
does not commute with amplitude damping noise in either the control or the target qubit, and it also does not commute with decoherence in the target qubit. 
It has also been shown experimentally that GI yields non-sensible results in noise-mitigated calibration procedure
\cite{npjqiKIK}. 
Although GI is widely used and is, in principle, compatible with any platform, this advantage is irrelevant if it produces incorrect results. 
Pulse-inverse (KIK) requires the ability to reverse the drive
sign in the effective Hamiltonian but yields correct mitigation.

\begin{figure}[h!]
\centering
\includegraphics[width=0.5\linewidth]{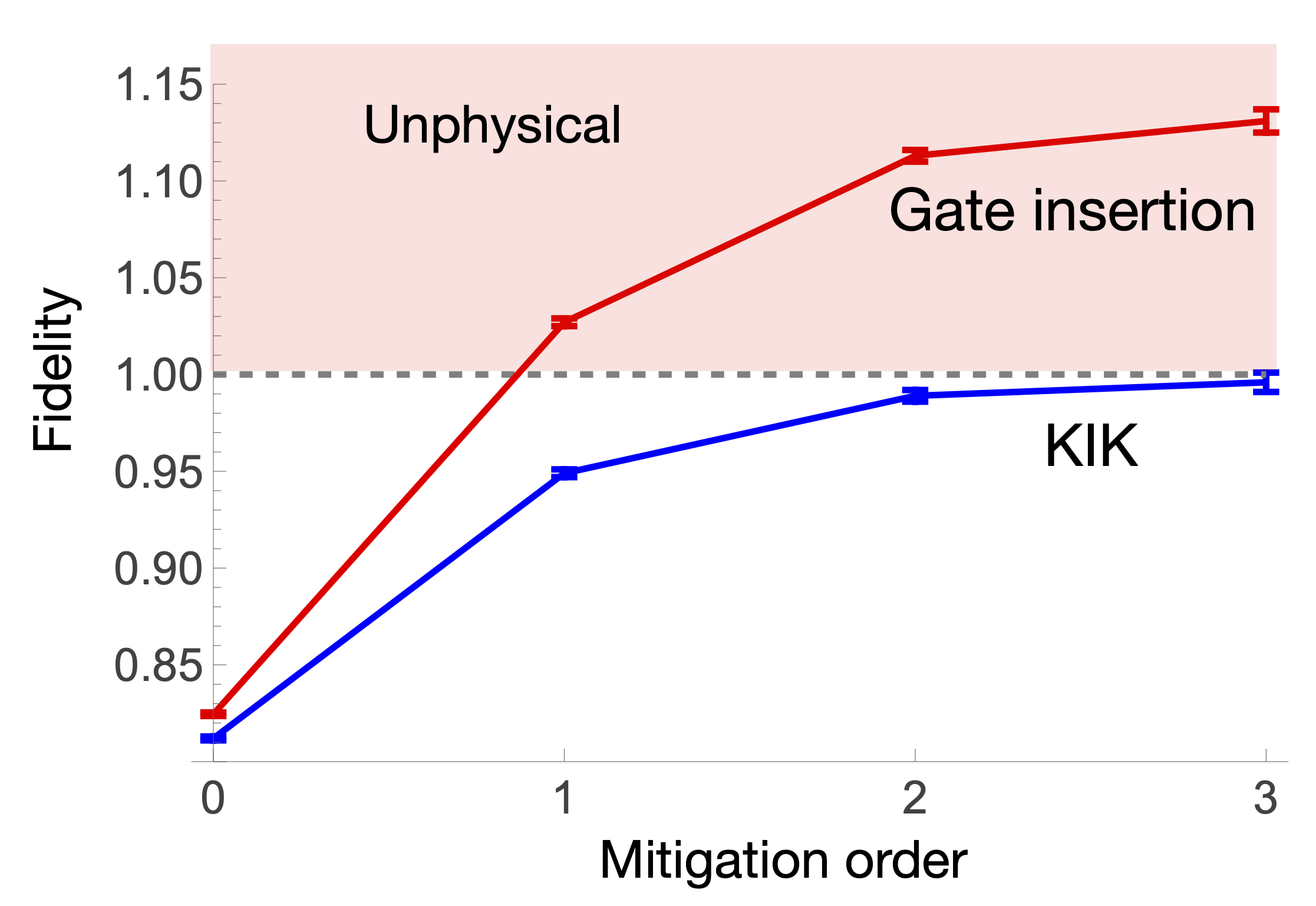}
\caption{\label{fig: GIPI}An experimental comparison of the fidelity of the mitigated quantum state, using gate insertion (red) and pulse-inverse KIK (blue) on the IBM processor \textit{ibm\_jakarta}. 
The circuit comprises of two qubits that undergo
ten swap operations.}
\end{figure}

To further evidence GI’s failure, we have performed a two-qubit
ten-swap experiment on \textit{ibm\_jakarta}. 
The initial state was set to $|11\rangle$, and $40,000$ shots were executed for each amplification circuit. 
We used qubit 0 and qubit 1. 
Supplementary Figure~\ref{fig: GIPI} (together with Table~\ref{tab:jakarta_data}) shows that gate insertion leads to unphysical values (fidelity larger than one) while the pulse-inverse (KIK) converges to fidelity 1. 
Although some observables or initial states may result in zero contribution from $[U,N]$, in general this term is non-zero and can lead to substantial errors, as demonstrated here.

\begin{table}[ht!]
\caption{\label{tab:jakarta_data} Experimental results from the IBM processor \textit{ibm\_jakarta}. The data presented in Tab.(a) and (b) are the same used to in Supplementary Figure~\ref{fig: GIPI}.}
\centering
\begin{minipage}[t]{0.45\textwidth}
\centering
{\small Tab. (a): Gate insertion data.}
\begin{tabular}{ccc}
\toprule
order & Unmitigated& mitigated value \\
\hline
0 & 0.8245(9) & 0.8245(9) \\
1 & 0.419(2)  & 1.027(2) \\
2 & 0.243(2)  & 1.113(3) \\
3 & 0.239(7)  & 1.131(6) \\
\hline
\end{tabular}
\end{minipage}
\hfill
\begin{minipage}[t]{0.45\textwidth}
\centering
{\small Tab. (b): KIK (pulse-inverse) data.}
\begin{tabular}{ccc}
\toprule
order & Unmitigated& mitigated value \\
\hline
0 & 0.812(1)  & 0.812(1) \\
1 & 0.538(2)  & 0.949(2) \\
2 & 0.370(1)  & 0.989(3) \\
3 & 0.284(3)  & 0.996(5) \\
\hline
\end{tabular}
\end{minipage}
\end{table}

\section{\hspace{-2mm}Using virtual Z gate to implement the pulse inverse in circuit level interface}
\label{SupplementaryNote3}

In various platforms, the single qubit Z gate is not implemented in practice. 
Instead, it is absorbed in the operation it preceded and propagated to the end of the circuit. 
Since Z gates only affect the phase in the computational basis, the Z gate before the measurement can be removed without affecting the measurement outcome.

Consider a gate $U$ (a single or multi-qubit gate) that does not commute with the single qubit partial $Z$ gate operation $Z(\theta)$, we can write
\begin{equation}
UZ(\theta)\to Z(\theta)Z(\theta)^{\dagger}UZ(\theta)=Z(\theta)U',
\end{equation}
where $U'$ is the gate $U$ dressed by the $Z(\theta)$ operation.
Although formally this transformation can always be done,
in some platforms, $U'$ is as easy to implement as $U$. 
For example, in the cross resonance gate, $U'$ differs from $U$ by the phase of the drive. 
When there are multiple gates, the same principle still holds
\begin{equation}
\sum_{i}\pi_{i}U_{3}U_{2}U_{1}Z(\theta)\to\sum_{i}\pi_{i}Z(\theta)U_{3}'U_{2}'U_{1}'=\sum_{i}\pi_{i}U_{3}'U_{2}'U_{1}',
\end{equation}
where $\pi_{i}$ is the projection operator of the computational state $i$ in Liouville space, so that $\sum_{i}\pi_{i}$ represents a measurement operation in the computation basis. Such implementation of a Z gate is referred to a virtual Z \cite{VirtZ}.

\subsection*{Cross resonance gate and Mølmer--Sørensen gate}

In the simple cross-resonance gate version without echo, to generate the $K_{I}$ we need to implement $R_{zx}(-\pi/2)$ instead of the $R_{zx}(+\pi/2)$. 
This can be achieved by using two virtual $Z$ gates $Z_{2}^{v}(\pi)$ on qubit number two, which is the target qubit associated with conditioned $x$ rotation.
\begin{equation}
Z_{2}^{v}(\pi)R_{zx}(+\pi/2)Z_{2}^{v}(\pi)=Z_{1}^{v}(\pi)Z_{1}^{v}(\pi)R_{zx}(-\pi/2)=R_{zx}(-\pi/2).
\end{equation}
The same principle holds for the Mølmer--Sørensen (MS) gate, where this time $Z^{v}(\pi)$ can be applied to either one of the $R_{xx}$ qubits.


\noindent\textit{The echo cross-resonance gate.}
For the echo version of the cross-resonance gate we have
\begin{equation}
R_{x,1}(\pi)R_{zx}(-\pi/4)R_{x,1}(\pi)R_{zx}(\pi/4),
\end{equation}
where $R_{x,1}(\pi)$ is an $X$ $\pi$-pulse on the first qubit.
The pulse inverse is
\begin{equation}
R_{zx}(-\pi/4)R_{x,1}(-\pi)R_{zx}(\pi/4)R_{x,1}(\pi).
\end{equation}
In the IBM cross-resonance processors (`Eagle'), the operation $R_{zx}(-\pi/4)R_{x,1}(\pi)R_{zx}(\pi/4)$ is called an ECR (echo cross resonance) gate. 
We point out that when it comes to noisy operation $R_{x,1}(\pi)\neq R_{x,1}(-\pi)$ as pointed out in \cite{Lidar2014DDreview,Lidar2024virtualZ}. 
These two operations are obtained using drive with opposite signs. 
In the ECR pulse, the $R_{zx}$ are already antisymmetric that the pulse inverse of an ECR pulse is
\begin{equation}
ECR_{I}=R_{zx}(-\pi/4)R_{x,1}(-\pi)R_{zx}(\pi/4).
\end{equation}
This sequence can be implemented by applying $Z_{1}^{v}(\pi)$ before and after the ECR,
\begin{align}
Z_{1}^{v}(\pi)ECRZ_{1}^{v}(\pi) & =R_{zx}(-\pi/4)Z_{1}^{v}(\pi)R_{x,1}(\pi)Z_{1}^{v}(\pi)R_{zx}(\pi/4)\nonumber \\
 & =R_{zx}(-\pi/4)R_{x,1}(-\pi)R_{zx}(\pi/4)=ECR_{I}.
\end{align}
We conclude that for the non-echo CR gate, the echo CR gate, and the MS gate, the pulse inverse can be implemented using a circuit interface by inserting virtual Z gates in the right place.

\section{\hspace{-2mm}$\Omega_{1}$ \& $\Omega_{2}$ Global vs Layer}
\label{SupplementaryNote4}

In the global form, a circuit composed of two layers a and b has the following evolution operator
\begin{equation}
K^{GKIK}=U_{b}U_{a}e^{\int_{0}^{t_{b}}L_{int}(t)dt+\frac{1}{2}\int_{0}^{t_{b}}dt'\int_{0}^{t}[L_{int}(t'),L_{int}(t)]dt}=U_{tot}e^{\Omega_{1}^{GKIK}+\Omega_{2}^{GKIK}},
\end{equation}
where
\begin{align}
\Omega_{1}^{GKIK} & =\int_{0}^{t_{b}}U^{\dagger}(t)L_{int}(t)U(t)dt\\
& =\int_{0}^{t_{a}}U^{\dagger}(t)L_{int}(t)U(t)dt+\int_{t_{a}}^{t_{b}}U^{\dagger}(t)L_{int}(t)U(t)dt\\
& \doteq\Omega_{1}^{a}+\Omega_{1}^{b}.\label{eq: omega1G}
\end{align}
$U(t)=U_{0\to t}$ refers to the noiseless evolution operator from
$t=0$ to time $t$. 
The second Magnus term is:
\begin{align}
\Omega_{2}^{GKIK} & =\frac{1}{2}\int_{0}^{t_{a}}dt'\int_{0}^{t'}[L_{int}(t'),L_{int}(t)]dt+\frac{1}{2}\int_{t_{a}}^{t_{b}}dt'\int_{t_{a}}^{t'}[L_{int}(t'),L_{int}(t)]dt\\
 & +\frac{1}{2}\int_{t_{a}}^{t_{b}}dt'\int_{0}^{t_{a}}[L_{int}(t'),L_{int}(t)]dt\circeq\Omega_{2}^{a}+\Omega_{2}^{b}+\frac{1}{2}[\Omega_{1}^{b},\Omega_{1}^{a}].\label{eq: omega2G}
\end{align}
Next, we repeat the same calculation from a layer point of view:
\begin{align}
K_{(2j+1)}^{LKIK}=U_{b}e^{\tilde{\Omega}_{1}^{b}+\tilde{\Omega}_{2}^{b}}U_{a}e^{\Omega_{1}^{a}+\Omega_{2}^{a}}  =
U_{b}U_{a}e^{U_{a}^{\dagger}(\tilde{\Omega}_{1}^{b}+\tilde{\Omega}_{2}^{b})U_{a}}e^{\Omega_{1}^{a}+\Omega_{2}^{a}},
\end{align}
where
\begin{align}
\tilde{\Omega}_{1}^{b} & =\int_{t_{a}}^{t_{b}}\tilde{L}_{int}(t)dt,\\
\tilde{L}_{int}(t) & =U_{t_{a}\to t}^{\dagger}L(t)U_{t_{a}\to t}.
\end{align}
Since
\begin{equation}
U_{t_{a}\to t}U_{a}=U(t),
\end{equation}
we get
\begin{equation}
\tilde{\Omega}_{1}^{b}=\int_{t_{a}}^{t_{b}}\tilde{L}_{int}(t)dt=U_{a}\left[\int_{t_{a}}^{t_{b}}U^{\dagger}(t)L(t)U(t)dt\right]U_{a}^{\dagger}=U_{a}\left[\int_{t_{a}}^{t_{b}}L_{int}(t)dt\right]U_{a}^{\dagger},
\end{equation}
and therefore 
\begin{equation}
U_{a}^{\dagger}\tilde{\Omega}_{1}^{b}U_{a}=U_{a}^{\dagger}\left[\int_{t_{a}}^{t_{b}}\tilde{L}_{int}(t)dt\right]U_{a}=\int_{t_{a}}^{t_{b}}L_{int}(t)dt.
\end{equation}
Since this holds for each one of the layers, we conclude that
\begin{equation}
K^{LKIK}=U_{b}U_{a}e^{\Omega_{1}^{b}+\Omega_{2}^{b}}e^{\Omega_{1}^{a}+\Omega_{2}^{a}}.
\end{equation}
 From the Baker-Campbell-Hausdorff (BCH) formula, we get
\begin{equation}
K^{LKIK}=U_{b}U_{a}e^{\Omega_{1}^{b}+\Omega_{2}^{b}}e^{\Omega_{1}^{a}+\Omega_{2}^{a}}\cong U_{b}U_{a}e^{(\Omega_{1}^{b}+\Omega_{1}^{a})+\{(\Omega_{2}^{b}+\Omega_{2}^{b})+\frac{1}{2}[\Omega_{1}^{b},\Omega_{1}^{a}]\}}.
\end{equation}
By comparing this result to Eq.~\eqref{eq: omega1G} and Eq.~\eqref{eq: omega2G} we conclude that in the no-noise amplification case $\Omega_{1}^{GKIK}=\Omega_{1}^{LKIK}$
and $\Omega_{2}^{GKIK}=\Omega_{2}^{LKIK}$. 
However, as explained in the main text, when applying the Taylor mitigation by combining different noise amplification factors, we find that the residual $\Omega_{2}$ in the case of LKIK mitigation is substantially and systematically smaller compared to the residual $\Omega_{2}$ in the GKIK case. 
In fact, for LKIK mitigation, in the limit of thin layers, the $\Omega_{2}$ term vanishes rapidly (scales as $1/L^{2}$ where $L$ is the number of layers).

\section{\hspace{-2mm}Properties of the Taylor Mitigation coefficients}
\label{SupplementaryNote5}

\subsection*{Sum of All Taylor Mitigation Coefficients}

First, we show the property
\begin{equation}
\label{sumone}
\sum_{j=0}^{M}a_{j}^{(M)}=1,
\end{equation}
where
\begin{equation}
a_{j}^{(M)}=\frac{(2M+1)!!(-1)^{j}}{2^{M}(M-j)!j!(2j+1)}.
\end{equation}
Using the binomial coefficients, we get
\begin{equation}
\sum_{j=0}^{M}a_{j}^{(M)}=
\sum_{j=0}^{M}\frac{(2M+1)!!(-1)^{j}}{2^{M}(M-j)!j!(2j+1)}
=\frac{(2M+1)!!}{2^{M+1}M!}
\sum_{j=0}^{M}\binom{M}{j}
\frac{(-1)^{j}}{j+1/2}.
\end{equation}
Next, we will make use of the identity \cite{graham1994concrete}
\begin{equation}
\sum_{r=0}^{M}\binom{n}{r}\frac{(-1)^{r}}{r+x}=\frac{1}{x}\frac{1}{\binom{n+x}{n}}
\end{equation}
which holds for $x\notin\{0,\pm1,\pm2,\dots\pm n\}$. For $n=M,\text{ }r=j,\text{ }x=\frac{1}{2}$
one has
\begin{equation}
\frac{(2M+1)!!}{2^{M+1}M!}\sum_{j=0}^{M}
\binom{M}{j}
\frac{(-1)^{j}}{j+\frac{1}{2}}
=\frac{(2M+1)!!}{2^{M+1}M!}
\frac{1}{\frac{1}{2}\binom{M+1/2}{M}}
=\frac{(2M+1)!!}{2^{M}M!}
\frac{1}{\binom{M+1/2}{M}}.
\label{sumtwo}
\end{equation}
In order to treat the choosing function of a half integer in the denominator of Eq.~\eqref{sumtwo}, one has to use the gamma function representation of the binomial coefficient:
\begin{equation}
\binom{x}{y}=\frac{\Gamma(x+1)}{\Gamma(y+1)\Gamma(x-y+1)}.
\end{equation}
Since for  an integer $M$, $\Gamma(M+1)=M!$ and for half integer
\begin{align}
\Gamma(1/2) & = \sqrt{\pi},\\
\Gamma(x+1/2) & = \frac{(2x-1)!! \sqrt{\pi}}{2^{x}},
\end{align}
we obtain 
\begin{align}
\binom{M+1/2}{M} = 
\frac{\Gamma(M+1+1/2)}{\Gamma(M+1)\Gamma(1/2+1)}
=\frac{(M+1/2)\Gamma(M+1/2)}{\frac{1}{2}M!\Gamma(1/2)}
\nonumber \\
=\frac{(2M+1)(2M-1)!!\sqrt{\pi}}{2^{M}\sqrt{\pi}M!}=\frac{(2M+1)!!}{2^{M}M!}.
\end{align}
Finally
\begin{equation}
\frac{(2M+1)!!}{2^{M}M!}\frac{1}{\binom{M+1/2}{M}}
=\frac{(2M+1)!!}{2^{M}M!}\frac{2^{M}M!}{(2M+1)!!}=1.
\end{equation}
$\blacksquare$

\subsection*{Vanishing moments}

Next, we derive the relation
\begin{equation}
\sum_{j=0}^{M}a_{j}^{(M)}\frac{(2j+1)^{l}}{k!}=0,
\label{sum3}
\end{equation}
for $k,l\leq M,l\neq0$. 
In order to show the wanted property (\ref{sum3}),
we use the identity \cite{graham1994concrete}
\begin{equation}
\sum_{r=0}^{n}\bigg[\ \binom{n}{r}(-1)^{r} Q(r)\bigg]=0,
\end{equation}
where $Q(r)$ is a polynomial in $r$ with a degree smaller than $n$.
Applying it to Eq.~\eqref{sum3}, we get
\begin{equation}
\sum_{j=0}^{M}\frac{(2M+1)!!(-1)^{j}(2j+1)^{l-1}}{2^{M}(M-m)!j!k!}=\frac{(2M+1)!!}{2^{M}k!M!}
\sum_{j=0}^{M}\binom{M}{j}(-1)^{j}(2j+1)^{l-1}=0,\label{sum4}
\end{equation}
since $(2j+1)^{l-1}$ is a polynomial in $j$ with a degree smaller
than $M$ for all $l$.\\
$\blacksquare$

\section{\label{SupplementaryNote6}\hspace{-2mm}Extension of the $\Omega_{2}$ bound to non-uniform layers}

Consider the expression $G=\sum_{i=1}^{N}\frac{1}{2}\Delta_{i}^{2}$ that appears in 
Eq.~(32) of the main text
where $\Delta_{i}=t_{i+1}-t_{i}$, and $\sum_{i=1}^{N}\Delta_{i}=\tau$.
Using a Lagrange multiplier, it is straightforward to show that $G$ is minimized for uniform-width layers. 
Starting with
\begin{equation}
G=\sum_{i=1}^N\frac{1}{2}\Delta_{i}^{2}+\lambda\bigg(\sum_{i=1}^N\Delta_{i}-\tau\bigg),
\end{equation}
and setting $\frac{dG}{d\Delta_{i}}=0$, yields
\begin{align}
\Delta_{i}+\lambda & =0.
\end{align}
Using this in $\sum_{i=1}^{N}\Delta_{i}=\tau$ yields $-N\lambda=\tau$, and therefore 
\begin{equation}
min(G)=\frac{1}{2}\frac{\tau^{2}}{N}.
\end{equation}
Next, we introduce stretch factors $\alpha_{i}$ such that the new
intervals are $\tilde{\Delta}_{i}=(\tau/N)(1+\alpha_{i})$, with $\sum\alpha_{i}=0$ for satisfying $\sum_{i=1}^{N}\tilde{\Delta}_{i}=\tau$. 
As a result, we get
\begin{align}
G=\sum\frac{1}{2}\tilde{\Delta}_{i}^{2} & =\sum_{i}\frac{1}{2}((\tau/N)(1+\alpha_{i}))^{2}=\frac{1}{2}\sum_{i=1}^{N}(\tau/N)^{2}(1+\alpha_{i}^{2})\nonumber \\
 & =\frac{\tau^{2}}{2N}+\frac{\tau^{2}}{2N}\frac{\sum_{i=1}^{N}\alpha_{i}^{2}}{N}.
\end{align}
As an illustrative example, if half of the intervals are 100\% longer, we obtain $\frac{3}{4}\frac{\tau^{2}}{N}$ instead of $\frac{1}{2}\frac{\tau^{2}}{N}$ for the uniform case.
The bound thus remains proportional to $1/N$.
Note that $\frac{\sum_{i=1}^{N}\alpha_{i}^{2}}{N}$ approximates the variance of $\alpha_{i}$ when $\alpha_{i}$ is sampled from a distribution, and becomes constant for large $N$.

\section{\hspace{-2mm}Comparison of the coefficient in \cite{Mari2024LayerBasedRich} and in the present work}
\label{SupplementaryNote7}

\begin{figure}[th!]
\includegraphics[width=0.6\linewidth]{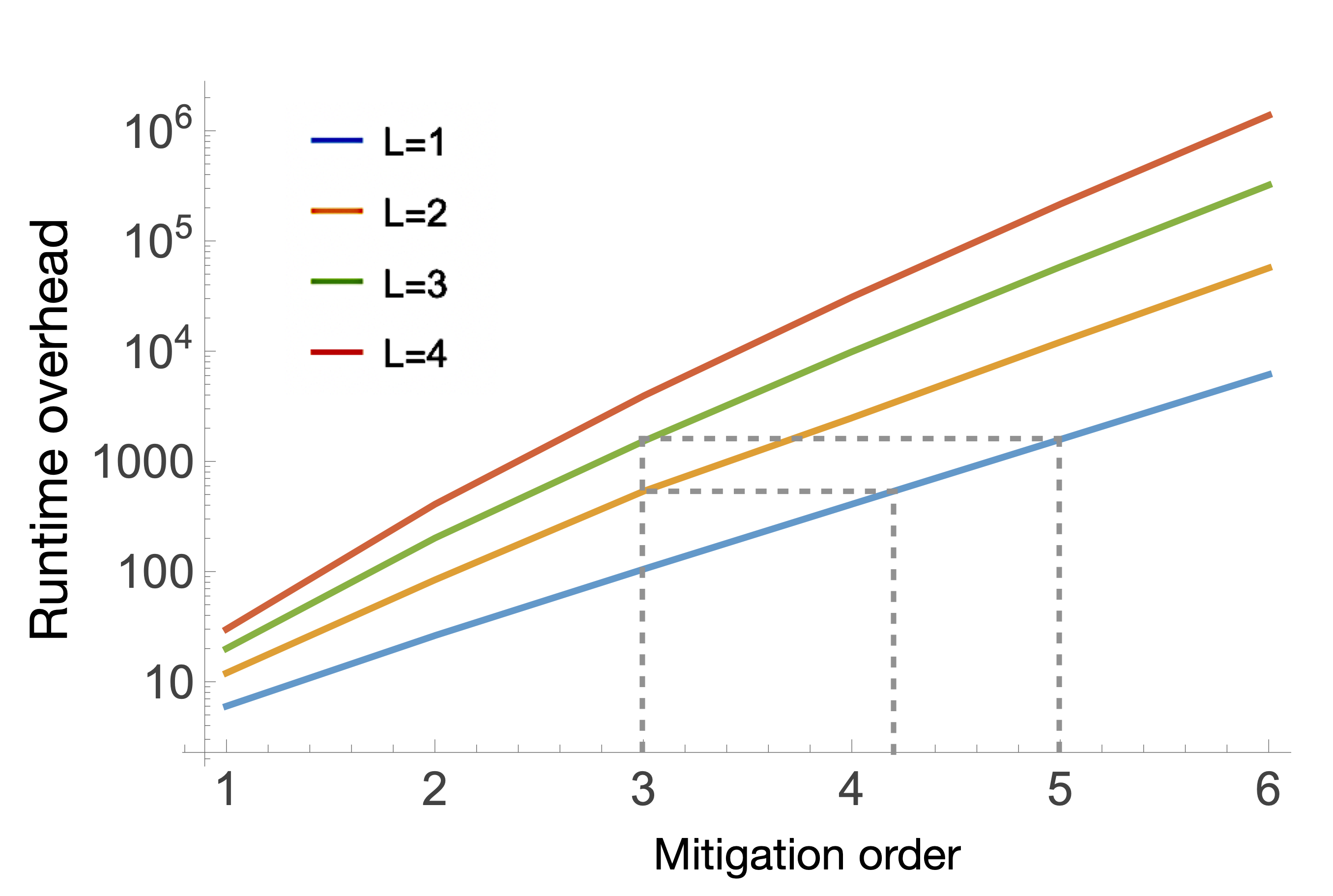}
\caption{\label{fig: MVE}A comparison of the total runtime cost $\gamma^{2}\langle depth\rangle$ of the multi-variate extrapolation (MVE) approach \cite{Mari2024LayerBasedRich}, as a function of the number of layers $L$. 
The mitigation order also describes the maximal noise power that is being suppressed, and is therefore an indicator of performance. 
Although then MVE circuits are, on average, shorter, the overall runtime cost is considerably higher than the SLT ($L=1$) used in the present paper. 
For example, the runtime overhead of third-order two-layer mitigation is not only considerably larger than that of a single layer at the same order, but is also comparable to fourth-order SLT mitigation (see dashed lines), which gives stronger mitigation at the same price. At the cost of using three layers, it is possible to execute fifth-order SLT mitigation.}
\end{figure}

In this Supplementary Note, we compare the runtime cost associated with the coefficients presented in \cite{Mari2024LayerBasedRich} 
(and in the section `The Layered KIK formula and its relation to other noise amplification schemes' of the present work),
which we term ``multi-variate extrapolation'' (MVE), to that of
a single layer Taylor (SLT) coefficients used in the present paper
(excluding section `The Layered KIK formula and its relation to other noise amplification schemes').
While MVE circuits are on average shorter than
SLT circuits, this advantage is minor compared to the considerable increase in sampling overhead. 
To quantify the runtime cost, we multiply the sampling overhead
\begin{equation}
\gamma^{2}=(\sum_{i}\left|c_{i}\right|)^{2},
\end{equation}
by the average depth
\begin{equation}
\left\langle depth\right\rangle =\frac{1}{\gamma}\sum_{i}\left|c_{i}\right|d_{i},
\end{equation}
where $d_{i}$ is the ratio between the depth of the noise-amplified circuit and the depth of the original circuit, and $c_{i}$ are the MVE or SLT coefficients. 
Supplementary Figure~\ref{fig: MVE} shows the runtime cost
$\gamma^{2}\left\langle depth\right\rangle $ as a function of the
mitigation order, for different numbers of layers. 
Note that MVE coefficients of a single layer are equal to the SLT coefficients. 
However, although we use SLT in the present paper, we do amplify the noise in layers using LKIK. 
If the amplification is done globally for a single layer,
the MVE will improve the accuracy, since it reduces the $\Omega_{2}$.
In this work we show that it is possible to remove the $\Omega_{2}$ without paying the heavy MVE runtime overhead.

As evident from Supplementary Figure~\ref{fig: MVE}, for a given order, increasing the number of layers increases the overhead, but should the performance of different number of layers be compared, for a given order? 
We claim that to a leading order, the answer is yes. 
When neglecting $\Omega_{2}$ effects, the mitigated evolution operator takes the form $K_{mit}^{(ord)}=U[I+O(\Omega_{1}^{ord+1})]$.
This holds for any layer number when using the MVE coefficients (including the SLT). 
While the pre-factors of the $O(\Omega_{1}^{ord+1})$ residual
terms depend on the layer number, the performance of all layers numbers will be dominated by $\Omega_{1}^{ord+1}$. 
Moreover, the dashed lines in Supplementary Figure~\ref{fig: MVE} show that for the same cost of the MVE (or even lower), it is possible to execute higher-order mitigation in SLT. 
\newpage


\end{document}